\documentclass{article}

\usepackage{arxiv}

\usepackage[utf8]{inputenc} 
\usepackage[T1]{fontenc}    
\usepackage{hyperref}       
\usepackage{url}            
\usepackage{booktabs}       
\usepackage{amsfonts}       
\usepackage{amsmath}
\usepackage{amssymb}
\usepackage{nicefrac}       
\usepackage{microtype}      
\usepackage{lipsum}		
\usepackage{graphicx}
\usepackage{natbib}
\usepackage{doi}
\usepackage{footmisc}
\usepackage{bbold}
\usepackage{listings}

\newcommand{\thetaB}{\boldsymbol\theta}
\newcommand{\piB}{\boldsymbol\pi}
\newcommand{\xiB}{\boldsymbol\xi}
\newcommand{\muB}{\boldsymbol\mu}
\newcommand{\UB}{\boldsymbol{U}}
\newcommand{\MB}{\boldsymbol{M}}
\newcommand{\DB}{\boldsymbol{D}}
\newcommand{\GB}{\boldsymbol{G}}
\newcommand{\cB}{\boldsymbol{c}}

\newcommand{\zeroB}{\boldsymbol{0}}

\usepackage{soul}
\usepackage{threeparttable}
\usepackage{bbm}
\usepackage{bbold}

\title{\LARGE\bf LORDs: Locally Optimal Restricted Designs for Phase I/II Dose-Finding Studies}

\author{ 
	\href{https://orcid.org/0000-0002-1626-2588}{\includegraphics[scale=0.06]{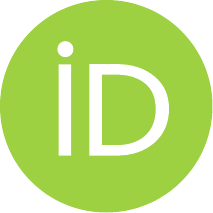}\hspace{1mm}Oleksandr Sverdlov} \\
	Novartis Pharmaceuticals Corporation\\
	East Hannover, NJ, USA\\
	\texttt{alex.sverdlov@novartis.com} \\    	
    \And
    \href{https://orcid.org/0000-0003-2997-8566}{\includegraphics[scale=0.06]{orcid.pdf}\hspace{1mm}Yevgen Ryeznik}\\
	Department of Mathematics\\
	Uppsala University\\
	Uppsala, Sweden \\
	\texttt{yevgen.ryeznik@math.uu.se} \\
    \And
    \href{https://orcid.org/0000-0001-5568-3054}{\includegraphics[scale=0.06]{orcid.pdf}\hspace{1mm}Weng Kee Wong} \\
	Department of Biostatistics \\
    University of California \\
    Los Angeles, CA, USA\\
	\texttt{wkwong@ucla.edu} \\	
}

\renewcommand{\shorttitle}{\textit{arXiv} Template}

\hypersetup{
pdftitle={A template for the arxiv style},
pdfsubject={q-bio.NC, q-bio.QM},
pdfauthor={David S.~Hippocampus, Elias D.~Striatum},
pdfkeywords={First keyword, Second keyword, More},
}

\begin{document}
\maketitle

\begin{abstract}
We propose Locally Optimal Restricted Designs (LORDs) for phase I/II dose-finding studies that focus on both efficacy and toxicity outcomes. As an illustrative
application, we find various LORDs for a 4-parameter continuation-ratio (CR) model defined on a user-specified dose range, where ethical constraints are imposed to prevent patients from receiving excessively toxic or ineffective doses. We study the structure and efficiency of LORDs across several experimental scenarios and assess the sensitivity of the results to changes in the design problem, such as adjusting the
dose range or redefining target doses. Additionally, we compare LORDs with a more heuristic phase I/II design and show that LORDs offer more statistically efficient and ethical benchmark designs. A key innovation in our work is the use of a nature-inspired
metaheuristic algorithm to determine dose-finding designs. This algorithm is free from assumptions, fast, and highly flexible. As a result, more realistic and adaptable designs for any model and design criterion with multiple practical constraints can be readily found and implemented. Our work also is the first to suggest how to modify and informatively select the next set of doses for the next study for enhanced statistical inference.
\end{abstract}


\section{Introduction}\label{Sec1}
The primary goal of a phase I oncology trial is to define the toxicity profile of the compound and establish the recommended dose for subsequent testing \citep{Ratain1993}. For phase~I oncology studies of cytotoxic compounds, this objective is usually equivalent to finding the maximum tolerated dose (MTD), defined as the highest dose at which the rate of toxicity is ``acceptable.'' An inherent assumption is that a higher dose, if it is not prohibitively toxic, is potentially more efficacious. Various phase~I dose escalation designs for identifying MTD have been proposed over the years; see, for example, \cite{Sverdlov2014} for an overview of these methods. 

Modern oncology drug development is increasingly focused on molecularly targeted therapies, immunotherapies, bispecific antibodies, and antibody drug conjugates  that challenge the assumption that MTD is the ``best'' dose. For instance, due to specific mechanisms of action, these novel treatment modalities can inhibit tumor growth or proliferation of cancer cells at doses below the MTD. Therefore, conventional phase~I toxicity-based dose escalation methods may not be the most appropriate as an integrated assessment of toxicity and efficacy (activity) may be essential to inform dose escalation. Such study designs are referred to as \emph{seamless phase I/II} designs \citep{Dragalin2010, Yan2018}, emphasizing the integration of the objectives of a standard phase~I trial (evaluating toxicity) and phase~II trial (evaluating early efficacy) in a single study. 

One important objective of a phase~I/II study is to ascertain the optimal biological dose (OBD), defined as ``the dose that has the highest desirability in terms of the risk-benefit trade-off'' \citep{Zhou2019}. To achieve this goal, various dose escalation designs based on efficacy and toxicity have been proposed, including nonparametric designs \citep{Ivanova2003}, parametric model-based designs such as the Eff-Tox method \citep{ThallCook2004}, model-assisted designs such as Bayesian optimal interval phase~I/II (BOIN12) design \citep{Lin2020}, among others.

The majority of phase~I and phase~I/II dose escalation designs are \emph{best intention designs}  \citep{Fedorov2011} in that they allocate each patient or cohorts of patients to the dose currently viewed as ``empirically best.'' 
This is done to protect the safety of study participants and comply with the ethical principles for clinical investigation \citep{WMA2013} which state that ``While the primary purpose of medical research is to generate new knowledge, this goal can never take precedence over the rights and interests of individual research subjects.'' However, the noble intention of maximizing individual patient benefit may in practice be compromised by poor learning about the best dose due to inherent limitations of early phase trials -- small sample sizes, heterogeneous patient populations, and high uncertainty on the compound's safety, tolerability, pharmacokinetics and pharmacodynamics. In contrast, \emph{optimal designs} are primarily concerned with maximizing the precision of the statistical inference \citep{Fedorov1972, FedorovLeonov2014}. 

When the sample size is large and the criterion is a convex function of the information matrices,   there is a unified mathematical approach to find model-based optimal designs and confirm their optimality.  They are theoretically most efficient among all designs  but they are rarely directly implemented in clinical oncology dose-finding trials.  This is because the optimal designs are derived mathematically and frequently may not appeal to practitioners.  For example, the optimal design has too few or too many doses than expected, or some doses are too close for  meaningful implementation.  More importantly, they may have potential conflicts with ethical principles, which were either not incorporated in the construction at the onset, or the optimization method cannot accommodate the constraints.   

The current paper proposes Locally Optimal Restricted Designs (henceforth referred to as ``LORDs'' or ``optimal designs'') in phase~I/II dose-finding trials. We hold the view that these designs are useful from multiple perspectives. First and foremost, they provide theoretical benchmarks for estimation precision against which other designs can be compared. Such a comparison can be very helpful at the study planning stage when the investigators are evaluating different options to find a design that fits the purpose of their study. In our paper, we focus on the continuation-ratio (CR) model \citep{FanChaloner2001, FanChaloner2004}, a widely used working model for phase~I/II trials with a trinomial outcome variable derived from bivariate binary efficacy--toxicity outcomes. Our ideas are generalizable to more complex models or design criteria. We build on the recent work of \cite{QiuWong2024} and use particle swarm optimization (PSO) for finding optimal designs with various restrictions that may be imposed by ethical imperatives (e.g., protecting study participants from overly toxic doses and doses that may have low potential for efficacy) and practical constraints (e.g., considering a discrete set of dose levels instead of the interval of doses).  Our designs are therefore an improvement over the previously published designs in \cite{FanChaloner2001, FanChaloner2004}, which required an unrestricted dose interval.  \cite{QiuWong2024}  did not incorporate ethical restrictions in the design stage, and consequently, the locally optimal designs   could have doses higher than the MTD.

Section~\ref{Sec2} provides statistical background for the CR model, formulates some important optimal design problems for a phase~I/II study, and describes how PSO can be used to efficiently find LORDs. Section~\ref{Sec3} investigates the structure of LORDs and their relative efficiency for selected experimental scenarios for the CR model. Section~\ref{Sec4} presents additional assessments of the sensitivity of LORDs to the choice of dose interval, the number of dose levels, and the definition of the targeted dose(s). We also provide an example of how LORDs can be compared in practice to a more heuristically chosen phase~I/II design. Section~\ref{Sec5} provides a discussion and outlines future work.

\section{Statistical background}\label{Sec2}

\subsection{The continuation-ratio model and the target doses}\label{Sec2.1}

Let $\mathfrak{X}=\{x_1,\ldots,x_K\}$ be a discrete set of pre-specified doses, where $x_1<\cdots<x_K$. For assessing dose--response, it is convenient to consider a continuous dose  interval $\mathfrak{D}=[d_L,d_U]$ such that $\mathfrak{X}\subset\mathfrak{D}$, where $d_L=x_1$ and $d_U=x_K$ are the lower and upper limits of the studied dose range. We call $d\in\mathfrak{D}$ the ``dose,'' recognizing in many cases $d=\ln(\text{dose})$. Assume that for each trial participant we have a bivariate binary outcome $(Y_T,Y_E)$:
\begin{equation*}
    Y_T=\begin{cases} 1,\; \text{if\;toxicity}; & \\  0,\; \text{otherwise.} \end{cases}
    \quad\quad
    Y_E=\begin{cases} 1,\; \text{if\;efficacy}; & \\  0,\; \text{otherwise}.\end{cases}
\end{equation*}

In many phase I/II trials, it is convenient to consider a trinomial outcome by defining a vector $\UB=(U_1, U_2, U_3)^\top$, with one entry equal to 1 and the other two entries equal to 0, depending on the combinations of efficacy and toxicity outcomes: $U_1=\mathbbm{1}\{Y_E=0, Y_T=0\}$ (no efficacy, no toxicity), $U_2=\mathbbm{1}\{Y_E=1, Y_T=0\}$ (efficacy without toxicity), and $U_3=\mathbbm{1}\{Y_T=1\}$ (toxicity). The outcome $\UB=(1,0,0)^\top$ is regarded as ``neutral,'' $\UB=(0,1,0)^\top$ as ``success,'' and $\UB=(0,0,1)^\top$ as ``toxicity.'' 

There are different ways to model the dose--response probabilities in the described setting. A useful approach is the CR model \citep{FanChaloner2001, FanChaloner2004}, which is a special case of ``contingent response'' models \citep{RabieFlournoy2013}. Consider two logistic regression models, one for the conditional probability of efficacy given no toxicity
\begin{equation}\label{logistic1}
p_{E|\bar{T}}(d)=\Pr(Y_E=1|Y_T=0,d)=\frac{e^{\theta_1+\theta_2d}}{1+e^{\theta_1+\theta_2d}},
\end{equation}
and the other one for the probability of toxicity
\begin{equation}\label{logistic2}
p_T(d)=\Pr(Y_T=1|d)=\frac{e^{\theta_3+\theta_4d}}{1+e^{\theta_3+\theta_4d}}.
\end{equation}Let $\thetaB=(\theta_1,\theta_2,\theta_3,\theta_4)^\top$, with the restrictions on the parameters: $\theta_1\ge \theta_3$, $\theta_3<0$, and $\theta_2,\theta_4>0$. Then the probabilities of ``neutral,'' ``successful,'' and ``toxic'' outcomes are expressed as 
\begin{eqnarray}
    \pi_0(d,\thetaB) &=& \Pr(\UB=(1,0,0)^\top|d) = (1-p_{E|\bar{T}}(d))(1-p_T(d))\nonumber \\
                     &=& \frac{1}{(1+e^{\theta_1+\theta_2d})(1+e^{\theta_3+\theta_4d})};\label{Pr_Z0}
\end{eqnarray}
\begin{eqnarray}
    \pi_1(d,\thetaB) &=& \Pr(\UB=(0,1,0)^\top|d) = p_{E|\bar{T}}(d)(1-p_T(d))\nonumber\\
                     &=& \frac{e^{\theta_1+\theta_2d}}{(1+e^{\theta_1+\theta_2d})(1+e^{\theta_3+\theta_4d})};\label{Pr_Z1}
\end{eqnarray}
\begin{eqnarray}
    \pi_2(d,\thetaB) &=& \Pr(\UB=(1,0,0)^\top|d) = p_T(d)=\frac{e^{\theta_3+\theta_4d}}{1+e^{\theta_3+\theta_4d}}.\label{Pr_Z2} 
\end{eqnarray} 

The outcome of a patient treated at dose $d\in\mathfrak{D}$ is $\UB\sim Multinomial(1,\piB(d))$, where $\piB(d)=(\pi_0(d,\thetaB),\pi_1(d,\thetaB),\pi_2(d,\thetaB))^\top$ with $\pi_0(d,\thetaB)+\pi_1(d,\thetaB)+\pi_2(d,\thetaB)=1$.

Several parameters may be of interest for the investigator. Their formal definition depends on whether the considered dose range (design space) is a continuous interval $\mathfrak{D}=[x_1,x_K]$ or a discrete set $\mathfrak{X}=\{x_1,\ldots,x_K\}$. 

Suppose the design space is a continuous interval $\mathfrak{D}=[x_1,x_K]$. The \emph{maximum tolerated dose} ($MTD$) is defined as a $100\Gamma$th percentile of the dose--toxicity curve, where $\Gamma\in(0,1)$ is a pre-specified target toxicity level. From~(\ref{Pr_Z2}), the $MTD$ can be expressed as 
\begin{equation}\label{MTD}
    MTD=\frac{\log(\frac{\Gamma}{1-\Gamma})-\theta_3}{\theta_4}.
\end{equation}While under the assumed logistic dose--toxicity model (\ref{logistic1}) the $MTD$ always exists and unique, it may fall outside $[x_1,x_K]$; e.g., it may be that $MTD>x_K$ if the target toxicity level is set ``too low.'' In practice, it is commonly assumed that the study doses have been determined appropriately and $MTD$ is properly contained in $[x_1,x_K]$.   

The \emph{optimal biological dose} ($OBD$) is defined as the dose that maximizes the probability of efficacy without toxicity, that is
\begin{equation}\label{OBD}
    OBD=\arg\max_{d}\pi_1(d,\thetaB).
\end{equation}
From (\ref{Pr_Z1}), $\pi_1(d,\thetaB)$ is maximized if and only if
$\theta_2(1+e^{\theta_3+\theta_4d}) - \theta_4(1+e^{-(\theta_1+\theta_2d)}) = 0$. The solution to this equation exists and is unique because $\theta_2(1+e^{\theta_3+\theta_4d})$ is monotonically increasing in $d$ and $\theta_4(1+e^{-(\theta_1+\theta_2d)})$ is monotonically decreasing in $d$. While $OBD$ can be regarded as the ``most desirable'' dose, it may fall outside $[x_1,x_K]$, and/or it may have higher than acceptable toxicity probability. This leads us to consider \emph{safe OBD}, which is defined by $OBD_s=\min(OBD, MTD)$ \citep{Ivanova2003}. The latter dose ($OBD_s$) could still fall outside the studied range of doses $[x_1,x_K]$, and $\pi_1(OBD_s,\thetaB)$ may be lower than some clinically desirable threshold value. However, for the purpose of the current paper, we omit these additional complexities and focus on $OBD$ as defined in (\ref{OBD}).

The \emph{minimum efficacious dose (MinED)} is often defined as the lowest dose that still has therapeutic value, i.e., any dose below $MinED$ would be viewed as lacking potential for efficacy. Following \cite{Whitehead2004}, we define $MinED$ as the dose for which the probability of a neutral outcome equals some pre-specified value $\Delta\in(0,1)$, that is:
\begin{equation}\label{MinED}
    MinED=\underset{d}{\arg}\{\pi_0(d,\thetaB)=\Delta\}.
\end{equation}Since (\ref{Pr_Z0}) is monotonically decreasing in $d$, $MinED$ exists and unique, but it could potentially fall outside $[x_1,x_K]$, e.g., it may be that $MinED<x_1$ if $\Delta$ is set ``too high.'' 

The interval $[MinED, MTD]$ is referred to as the ``therapeutic window'' ($TW$). Doses within $TW$ merit further consideration. It can be that $TW=\varnothing$, in which case no dose satisfies the desired quality objectives. In the current paper, we do not address the questions of under which conditions on the CR model parameters $\thetaB$, the chosen thresholds $\Gamma$ and $\Delta$, and the chosen interval of doses $[x_1,x_K]$, the target doses (\ref{MTD}), (\ref{OBD}), and $(\ref{MinED})$ are properly contained in $[x_1,x_K]$ and $TW$ is nonempty. These questions will be addressed elsewhere. Instead, we focus on optimal designs for efficient estimation of these doses and acknowledge that they may fall outside $[x_1,x_K]$.

Suppose the design space is a discrete set $\mathfrak{X}=\{x_1,\ldots,x_K\}$. In this case, the target doses must be defined in terms of the doses from $\mathfrak{X}$. Following \cite{Cheung2011}, we define discrete analogues of (\ref{MTD})--(\ref{MinED}) as 
\begin{eqnarray}
    MTD^\prime &=& \arg\min_{1\le i\le K}|\pi_2(x_i,\thetaB)-\Gamma|;\label{MTD.prime}\\
    OBD^\prime &=& \arg\max_{1\le i\le K}\pi_1(x_i,\thetaB);\label{OBD.prime}\\
    MinED^\prime &=& \arg\min_{1\le i\le K}|\pi_0(x_i,\thetaB)-\Delta|.\label{MinED.prime}
\end{eqnarray}Depending on the chosen $\mathfrak{X}$, ``discrete'' target doses (\ref{MTD.prime})--(\ref{MinED.prime}) may be quite different from the ``continuous'' target doses (\ref{MTD})--(\ref{MinED}). In this paper, we do not address the questions related to the optimal choice of $\mathfrak{X}$ but rather focus on finding optimal designs for efficient estimation of doses (\ref{MTD.prime})--(\ref{MinED.prime}) assuming $\mathfrak{X}$ is provided by the investigator.

\subsection{The Fisher information matrix and optimal design problems}\label{Sec2.2}
Throughout, we assume we have a pre-determined sample size $n$ and $n_1,\ldots,n_K$ patients treated at the doses $x_1,\ldots,x_K$, respectively, such that $n_1+\ldots+n_K=n$. Let $\UB_{ij}$ be the response vector of the $j$th patient at the $i$th dose ($i=1,\ldots,K$, $j=1,\ldots,n_i$). Then the maximum likelihood estimate (MLE) of $\thetaB$ can be obtained as a solution to the system of score equations \citep{QiuWong2024}:
\begin{equation}
    \widehat{\thetaB} = \underset{\thetaB}{\arg}
    \left\{\sum_{i=1}^K\sum_{j=1}^{n_i}\GB(x_i)^\top\DB(x_i,\thetaB)\UB_{ij}=\zeroB\right\},
\end{equation}where
\begin{equation*}
    \GB(x_i)=\begin{pmatrix}
        1 & x_i & 0 & 0 \\
        0 & 0 & 1 & x_i \\
        0 & 0 & 0 & 0
    \end{pmatrix}
    \quad\text{and}\quad
    \DB(x_i,\thetaB)=\begin{pmatrix}
        \pi_0^{-1}(x_i,\thetaB) & 0 & 0 \\
        0 & \pi_1^{-1}(x_i,\thetaB) & 0 \\
        0 & 0 & \pi_2^{-1}(x_i,\thetaB) 
    \end{pmatrix}.
\end{equation*}The information matrix at dose $x_i$ is 
\begin{equation}\label{FIM_single}
    \muB(x_i,\thetaB) = 
    \GB(x_i)^\top\DB(x_i,\thetaB)\GB(x_i), 
\end{equation}and the Fisher information matrix (FIM) for $\thetaB$ from the sample of $n$ patients is $$\MB_n(\thetaB)=\sum_{i=1}^Kn_i\muB(x_i,\thetaB).$$

For an optimal design problem on the interval of doses $\mathfrak{D}=[x_1,x_K]$, one can consider a continuous design measure $\xiB=\{(d_i,\rho_i),i=1,\ldots,N\}$, where $\rho_i\in(0,1)$ is the allocation proportion for $d_i$ with $\sum_{i=1}^N\rho_i=1$. Here the set of dose levels $d_i\in\mathfrak{D}$ ($i=1,\ldots,N$) may be different from the pre-specified set of doses $x_i\in\mathfrak{X}$ ($i=1,\ldots,K$). The worth of a design $\xiB$ is measured by the FIM, $\MB(\xiB,\thetaB)=\sum_{i=1}^N\rho_i\muB(d_i,\thetaB)$, and $\MB^{-1}(\xiB,\thetaB)$ provides the lower bound on the variance-covariance matrix of an efficient estimator of $\thetaB$. Two ``benchmark'' designs are the D-optimal design minimizing the volume of the confidence ellipsoid for $\thetaB$,
\begin{equation}
    \xiB_D=\arg\min_{\xiB}\{-\ln|\MB(\xiB,\thetaB)|\},
\end{equation} and the c-optimal design for minimizing the asymptotic variance of a given estimated function of $\thetaB$.  In our case, the interest is in estimating $OBD$, and the sought c-optimal design is defined by
\begin{equation}
    \xiB_c=\arg\min_{\xiB}\{\cB^\top(\thetaB)\MB^{-1}(\xiB,\thetaB)\cB(\thetaB)\},
\end{equation}where $\cB(\thetaB)=\frac{\partial}{\partial\thetaB}OBD(\thetaB)$. 

One potential limitation of designs $\xiB_D$ and $\xiB_c$ is that they can have some allocation to doses that are too toxic and/or lacking efficacy. To address this limitation, we consider four \emph{restricted} optimal design problems:

\begin{itemize}
    \item[I.] A design minimizing $-\ln|\MB(\xiB,\thetaB)|$ under the restriction that the dose levels chosen from $\mathfrak{D}$ should not exceed $MTD$. 
    \item[II.] A design minimizing $-\ln|\MB(\xiB,\thetaB)|$ under the restriction that the dose levels chosen from $\mathfrak{D}$ should be within therapeutic window $[MinED, MTD]$. 
    \item[III.] A design minimizing asymptotic variance of the efficient estimator of $OBD$ under the restriction that the dose levels chosen from $\mathfrak{D}$ should not exceed $MTD$. 
    \item[IV.] A design minimizing asymptotic variance of the efficient estimator of $OBD$ under the restriction that the dose levels chosen from $\mathfrak{D}$ should be within therapeutic window $[MinED, MTD]$.
\end{itemize} 

For an optimal design problem on the discrete set of doses 
$\mathfrak{X}=\{x_1,\ldots,x_K\}$, we have a design measure 
$\xiB^\prime=\{(x_i,\rho_i^\prime),\;i=1,\ldots,K\}$ with $x_i\in\mathfrak{X}$, $\rho_i^\prime\in[0,1]$, and $\sum_{i=1}^K\rho_i^\prime=1$. The design FIM is $\MB(\xiB^\prime,\thetaB)=\sum_{i=1}^N\rho_i^\prime\muB(x_i,\thetaB)$. In this setup, we consider four restricted optimal design problems that are discrete analogues of I -- IV:
\begin{itemize}
    \item[I$^\prime$.] A design minimizing $-\ln|\MB(\xiB^\prime,\thetaB)|$ under the restriction that the dose levels chosen from $\mathfrak{X}$ should not exceed $MTD^\prime$. 
    \item[II$^\prime$.] A design minimizing $-\ln|\MB(\xiB^\prime,\thetaB)|$ under the restriction that the dose levels chosen from $\mathfrak{X}$  should be within therapeutic window $[MinED^\prime, MTD^\prime]$. 
    \item[III$^\prime$.] A design minimizing asymptotic variance of the efficient estimator of $OBD^\prime$ under the restriction that the dose levels chosen from $\mathfrak{X}$ should not exceed $MTD^\prime$. 
    \item[IV$^\prime$.] A design minimizing asymptotic variance of the efficient estimator of $OBD^\prime$ under the restriction that the dose levels chosen from $\mathfrak{X}$ should be within therapeutic window $[MinED^\prime, MTD^\prime]$.
\end{itemize} 

The stated optimal design problems are well-defined and can be solved using convex optimization methods. The design optimality can be checked using the General Equivalence Theorem (GET) \citep{KieferWolfowitz1960}, by plotting the sensitivity function (directional derivative) of a chosen design criterion and verifying that the derivative is zero at the design points and is negative otherwise.

\subsection{Finding LORDs using Particle Swarm Optimization}\label{subsec:pso}

To find an optimal design structure, one has to search for an extremum of the objective function representing the D- or c-optimality criterion, in the space of discrete probability measures. We use a Particle Swarm Optimization (PSO) algorithm \citep{pso1995, Luke2013Metaheuristics} to find LORDs. PSO is an iterative, gradient-free, stochastic optimization approach inspired by the collective behavior of animals such as ants or bees when searching for food. It operates in a multidimensional space of real-valued vectors, where vector coordinates represent the position of a particle in the search space, i.e., if $f(\mathbf{x}): \mathcal{X}\rightarrow \mathbb{R}$ is an objective function with $\mathbf{x}=\left(x_1, \ldots, x_n\right)^\top \in \mathcal{X}\subseteq \mathbb{R}^n$, then the position of a particle $P$ is given by $(x_1, \ldots, x_n)$. 

For a given objective function $f(P)$, the algorithm involves a series of iterations to move the swarm to the optimum point. The algorithm has several user-specified parameters, including the swarm size ($S$), the maximum number of iterations ($N_{\max}$), an inertia weight ($w$), a cognitive coefficient ($c_1$), and a social coefficient ($c_2$). A detailed description of the PSO algorithm is provided in the Appendix.

For finding LORDS on a continuous dose set $\mathfrak{D}$, we use the following transformation of a design $\xiB$ to a particle $P$:

\begin{equation}\label{PSO.transform}
    \xiB = \begin{pmatrix}
    d_1, \ldots, d_n \\
    \rho_1, \ldots, \rho_n
\end{pmatrix} \quad\Rightarrow\quad
P = \left(p_1, \ldots, p_n, p_{n+1}, \ldots, p_{2n}\right),
\end{equation}where
\begin{equation}\label{PSO.elements}
    p_1 = d_1, \;\ldots,\; p_n = d_n, p_{n+1} = \rho_1, \;\ldots,\; p_{2n} = \rho_n.
\end{equation}In (\ref{PSO.transform}), $n$ is initially set to $n=\frac{4\times 5}{2}=10$ (which is the upper bound to the number of support points of a D-optimal design for a 4-parameter model), but during the search for optimum, the number of support points may be reduced, depending on the underlying dose--response and the considered optimization problem. In (\ref{PSO.elements}), $p_1,\ldots,p_n$ are restricted by the considered dose interval, and $p_{n+1},\ldots,p_{2n}$ correspond to the allocation proportions that are nonnegative and must sum to one. To ensure the latter condition holds, the following transformation is performed for each particle at every iteration step:
$$
\rho_i = p_{n+i} := \frac{p_{n+i}^2}{\sum_{i=1}^np_{n+i}^2}, \: i = 1, \ldots, n.
$$

For finding LORDs on a discrete set of doses $\mathfrak{X}$, the design points are given as $x_1,\ldots,x_K$, and the optimization problem boils down to determining the allocation proportions, $\rho_1^\prime,\ldots,\rho_K^\prime$, under the constraints $\rho_i^\prime\in[0,1]$ and $\sum_{i=1}^K\rho_i^\prime=1$, that maximize the selected criterion.  

In all our computations, we use the following setup of the PSO algorithm:
$$
S=50, \: N_{\max} = 1500, \: c_1 = 2.5, \:c_2 = 0.5.
$$The \text{inertia weight} varies with iterations, starting at 0.9 and decreasing up to 0.4 (if $N_{\max}$ number of iterations is performed):
$$
w_j = 0.4 + 0.5\left(\frac{N_{\max}-j}{N_{\max}-1}\right)^\gamma,
$$
where $\gamma$ is a \textit{relaxation} parameter, and it is set to 1.25.

\section{The structure and efficiency of LORDs}\label{Sec3}

In the current paper, and similar to the earlier work by \cite{QiuWong2024}, we use PSO to find LORDs. For illustrative purposes, we consider four scenarios of CR dose--response, inspired by \cite{Whitehead2004}, summarized in Table~\ref{Table1}. For all scenarios, we assume $d=\ln(\text{dose})$, and the set $\mathfrak{X}$ contains nine doses $x_1<\cdots<x_9$, which are on the log scale
$$-1.20,\; -0.23,\; 0.92,\; 2.02,\; 3.00,\; 3.69,\; 4.38,\; 5.08,\;\text{and}\;5.77.$$Therefore, $\mathfrak{D}=[x_1,x_9]=$ [-1.20, 5.77], which corresponds to a range of doses 0.3~mg to 320~mg on the original (untransformed) scale. We also set $\Gamma=0.2$ and $\Delta=0.2$. 

\begin{figure}
    \centering
    \includegraphics[width=0.95\linewidth]{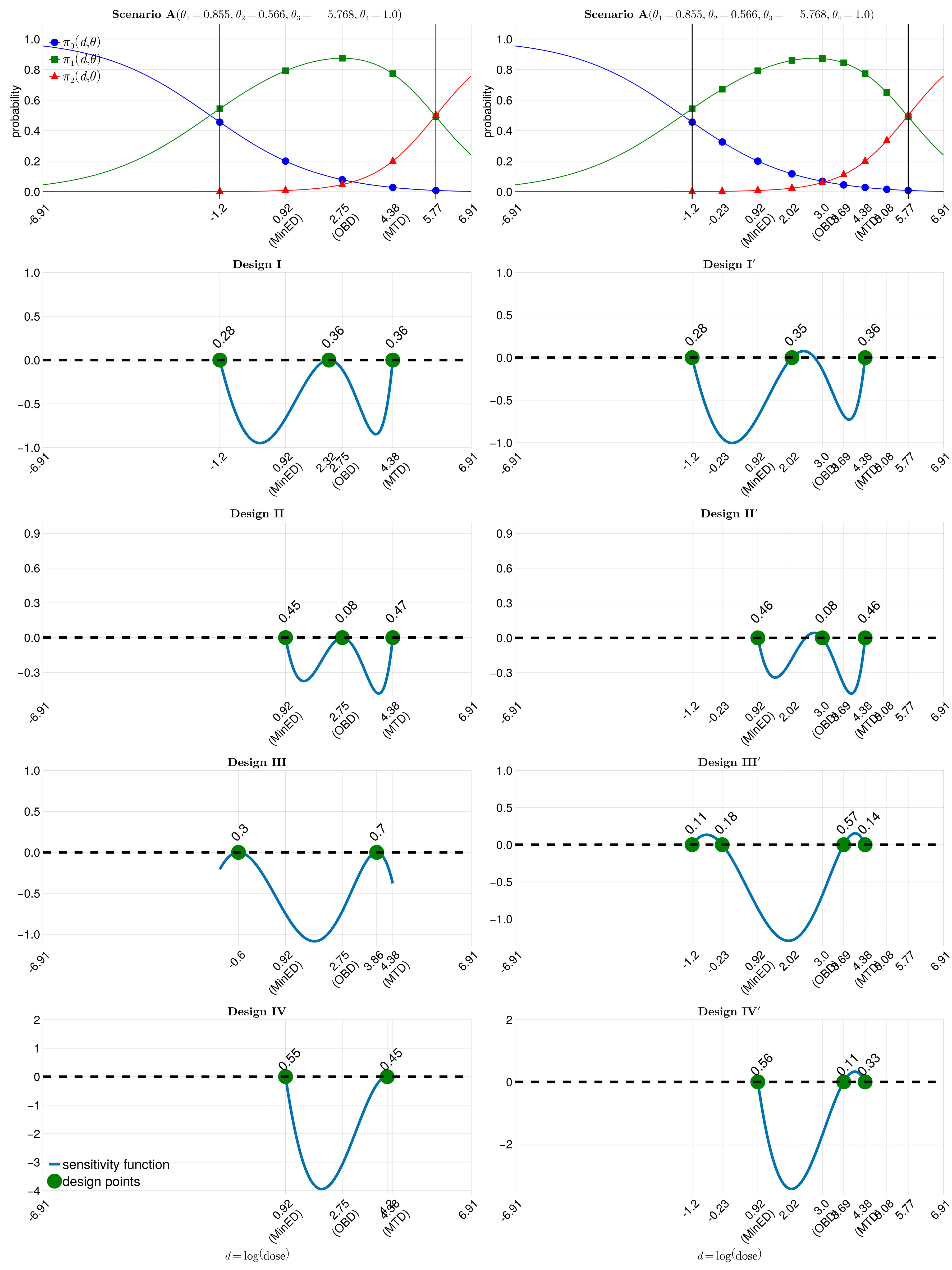}
    \caption{Optimal designs for scenario A (cf. Section~\ref{Sec3}).}
    \label{fig:scenarioA}
\end{figure}

\begin{figure}
    \centering
    \includegraphics[width=0.95\linewidth]{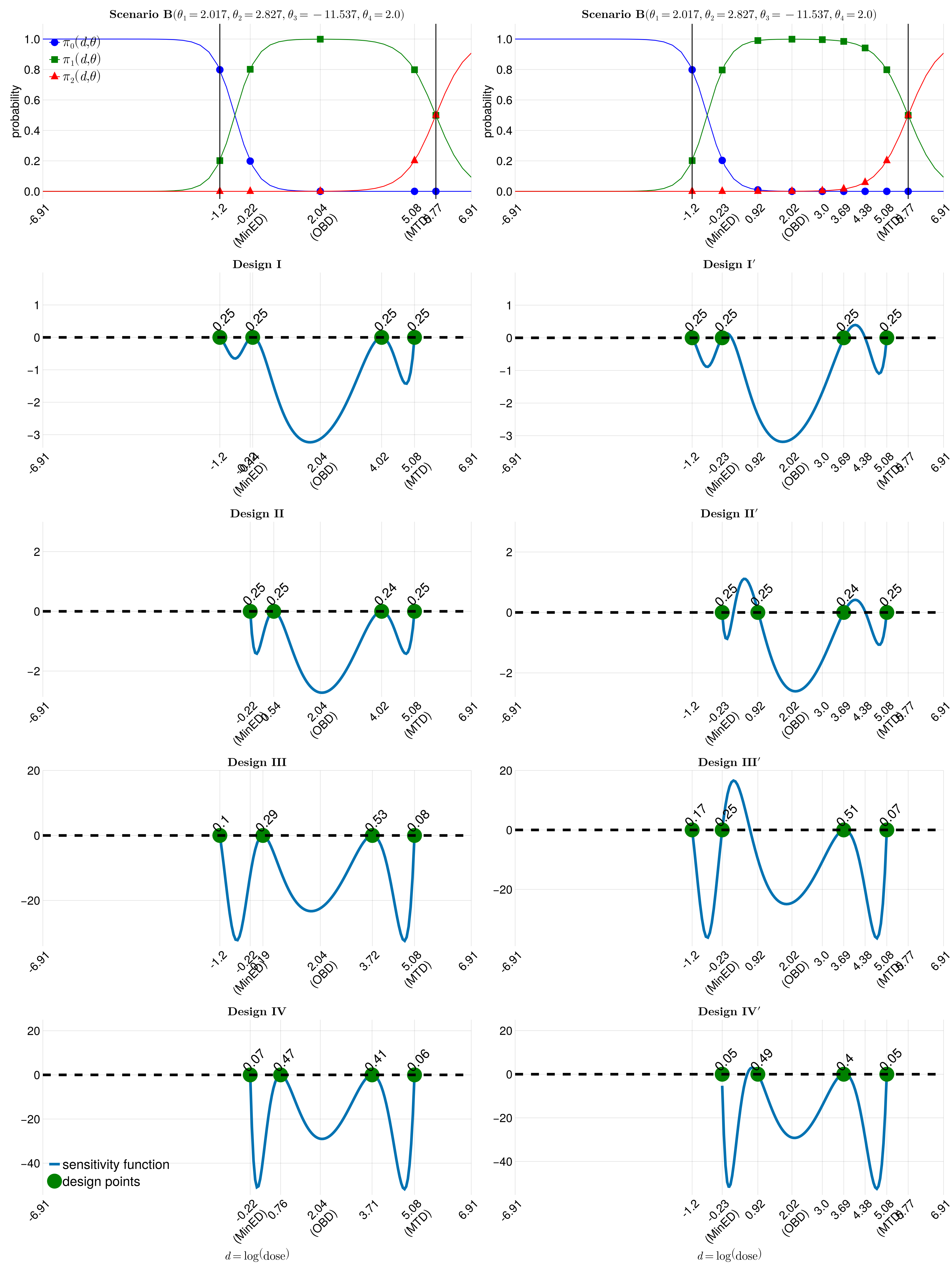}
    \caption{Optimal designs for scenario B (cf. Section~\ref{Sec3}).}
    \label{fig:scenarioB}
\end{figure}

\begin{figure}
    \centering
    \includegraphics[width=0.95\linewidth]{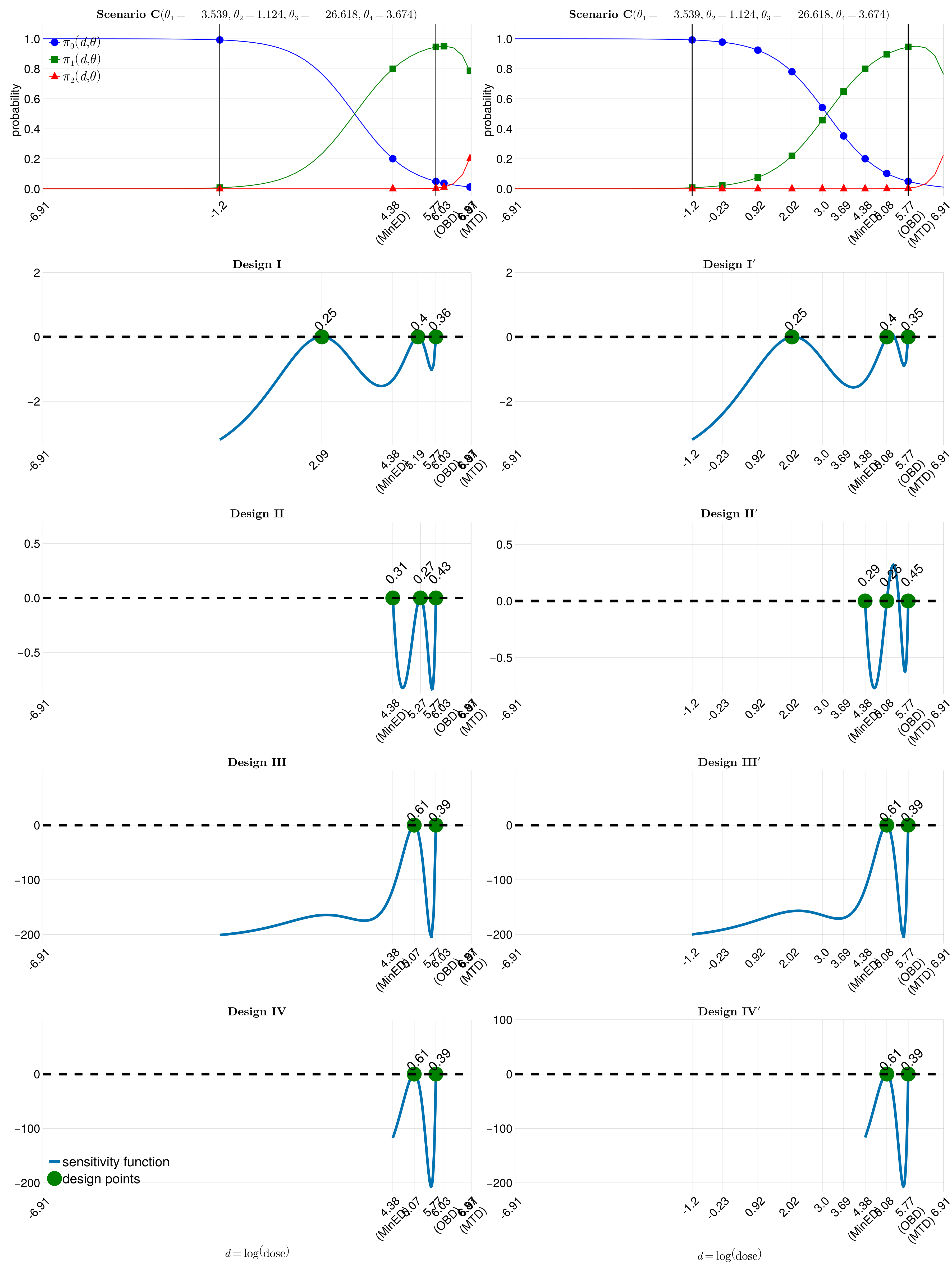}
    \caption{Optimal designs for scenario C (cf. Section~\ref{Sec3}).}
    \label{fig:scenarioC}
\end{figure}

\begin{figure}
    \centering
    \includegraphics[width=0.95\linewidth]{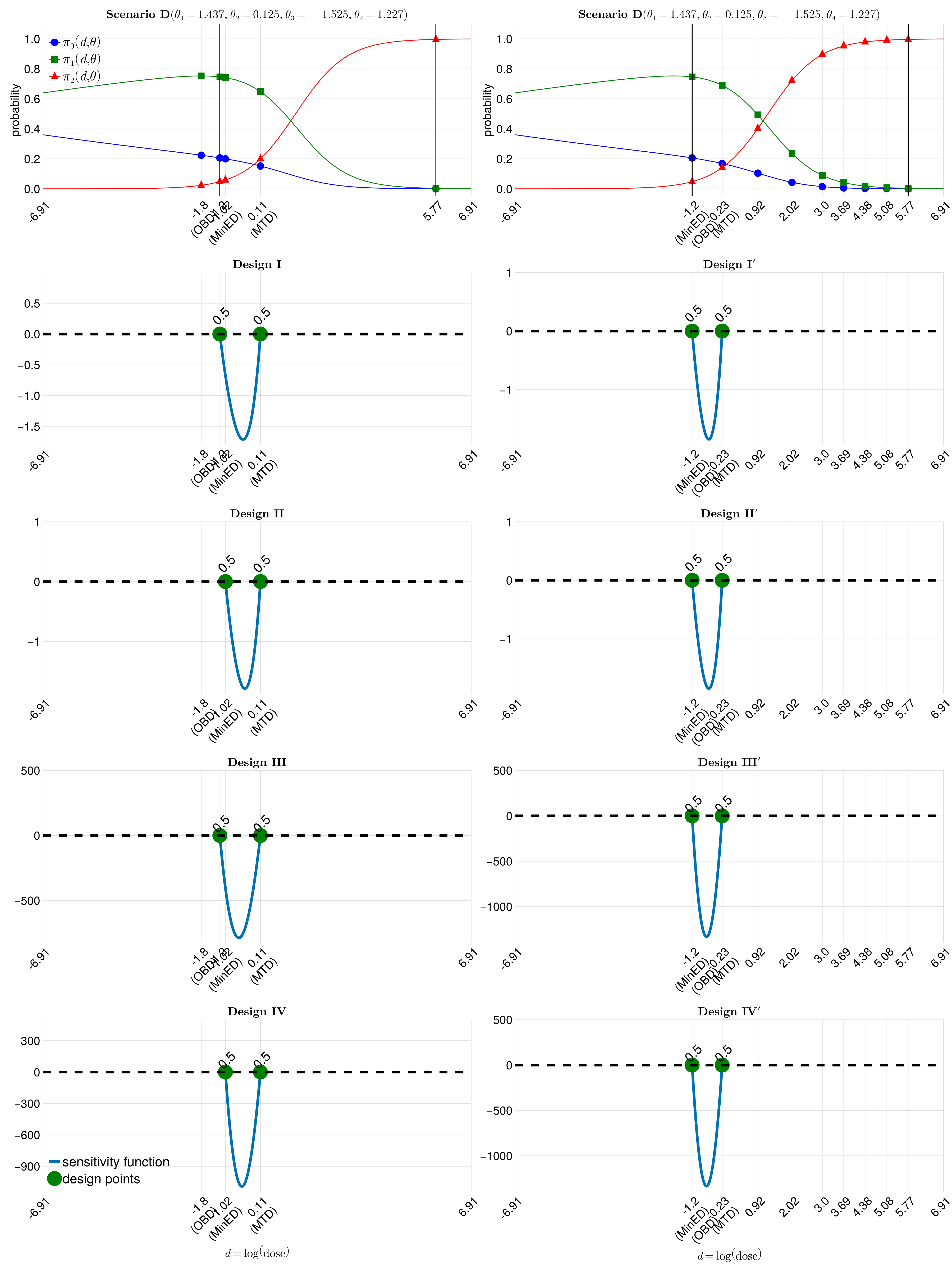}
    \caption{Optimal designs for scenario D (cf. Section~\ref{Sec3}).}
    \label{fig:scenarioD}
\end{figure}

The four dose--response scenarios, henceforth referred to as A, B, C, and D, are displayed in Figures~\ref{fig:scenarioA}, \ref{fig:scenarioB}, \ref{fig:scenarioC}, and~\ref{fig:scenarioD}, respectively, as continuous functions of dose (top left plots) and as discrete functions on the lattice of doses from $\mathfrak{X}$ (top right plots). Scenario~A (``Narrow'') corresponds to a case when $TW=[MinED, MTD]=[2.5,80]$, which is relatively narrow compared to the studied dose range $[0.3, 320]$.  Scenario~B (``Wide'') corresponds to a much wide $TW=[MinED, MTD]=[0.8,160]$. Scenario~C (``Safe'') represents a case when all study doses are safe, the toxicity probability at the highest dose is $0.00438<0.2$, and $MTD=963$~mg, which is above the upper limit of the dose range (320~mg). Scenario~D (``Unsafe'') represents a case when the toxicity is increasing at very low doses and $TW=[MinED,MTD]=[0.36,1.12]$. In Scenario~D, $OBD=0.2$~mg, which is below the lower limit of the dose range (0.4~mg). 


\begin{table*}
    \begin{threeparttable}
    \caption{Dose--response scenarios (A, B, C, D) of the CR model with $d=\ln(\text{dose})\in\mathfrak{D}$ = [-1.20, 5.77]. The thresholds for minimum efficacy and maximum toxicity probabilities are set to $\Delta=0.2$ and $\Gamma=0.2$.} 
    \centering
    \begin{tabular}{p{4cm}p{2.5cm}p{2.5cm}p{2.5cm}p{3cm}}
    \hline
     & \multicolumn{4}{c}{Scenario} \\
    \cline{2-5}     
    Parameters  & A & B & C & D \\
    \hline 
    $\theta_1$ &  0.855 &   2.017 &   -3.539 &  1.437 \\
    $\theta_2$ &  0.566 &   2.827 &    1.124 &  0.125 \\
    $\theta_3$ & -5.768 & -11.537 &  -26.618 & -1.525 \\
    $\theta_4$ &    1.0 &     2.0 &    3.674 &  1.227 \\
    \hline
    \multicolumn{5}{c}{Continuous dose space $\mathfrak{D}= $ [-1.20, 5.77] (on the original scale 0.3~mg to 320~mg)}\\
    \hline
    $\ln(MinED)$ & 0.92 & -0.22 & 4.38 & -1.02 \\
    $\ln(OBD)$ & 2.75 & 2.04 & 6.03 & -1.80$^*$ \\
    $\ln(MTD)$ & 4.38 & 5.08 & 6.87$^{**}$ & 0.11 \\
    $[MinED,MTD]$ & [2.5, 80] & [0.8, 160] & [80, 963] & [0.4, 1.1] \\
    $OBD$ & 15.6 & 7.7 & 417 & 0.2 \\
    \hline
    \multicolumn{5}{c}{Discrete dose space $\mathfrak{X}=\{x_1,\ldots,x_9\}^{***}$}\\
    \hline
    $\ln(MinED^\prime)$ & $x_3=0.92$ & $x_2=$ -0.23 & $x_7=4.38$ & $x_1= $ -1.20 \\
    $\ln(OBD^\prime)$ & $x_5=3.00$ & $x_4=2.02$ & $x_9=5.77$ & $x_1= $ -1.20 \\
    $\ln(MTD^\prime)$ & $x_7=4.38$ & $x_8=5.08$ & $x_9=5.77$ & $x_2= $ -0.23 \\
    $[MinED^\prime,MTD^\prime]$ & [2.5, 80] & [0.8, 160] &  [80, 320] &  [0.3, 0.8] \\
    $OBD^\prime$ & 20.0 & 7.5  & 320 &  0.3 \\
    \hline
    \end{tabular}\label{Table1}
    \begin{tablenotes}
    \footnotesize
       \item[*] $\ln(OBD)=$ -1.80 is below -1.20, the lower limit of the studied dose range.  
       \item[**] $\ln(MTD)=6.87$ is above 5.77, the upper limit of the studied dose range. The toxicity probability at the highest dose is $\pi_2(5.77,\thetaB)=0.0043$, which is much lower than the target toxicity level $\Gamma=0.2$.
       \item[***] $\mathfrak{X}$ contains nine doses: $x_1= $ -1.20, $x_2= $ -0.23, $x_3=0.92$, $x_4=2.02$, $x_5=3.00$, $x_6=3.69$, $x_7=4.38$, $x_8=5.08$, and $x_9=5.77$. 
    \end{tablenotes}    
    \end{threeparttable}
\end{table*}

Table~\ref{Table2} shows the structure of LORDs, and Table~\ref{Table3} presents the D- and c-efficiencies of these designs under the four scenarios of CR dose--response. The D-efficiency of a design $\xiB$ is computed relative to design~I, and it is defined as
\begin{equation}\label{D_efficiency}
    D_{\text{eff}}(\xiB,\thetaB) = \left\{\frac{|\MB^{-1}(\text{I},\thetaB)|}
    {|\MB^{-1}(\xiB,\thetaB)|}\right\}^{1/4}.
\end{equation}The c-efficiency of design $\xiB$ is computed relative to design~III, and it is defined as 
\begin{equation}\label{c_efficiency}
    c_{\text{eff}}(\xiB,\thetaB) = 
    \frac{\cB^\top(\thetaB)\MB^{-1}(\text{III},\thetaB)\cB(\thetaB)}
    {\cB^\top(\thetaB)\MB^{-1}(\xiB,\thetaB)\cB(\thetaB)}.
\end{equation}For instance, $D_{\text{eff}}(\xiB,\thetaB)=0.90$ means that the design $\xiB$ is 90\% as efficient as design~I, and the sample size for a study with design $\xiB$ must be increased by 10\% to achieve the same level of estimation efficiency as with design~I.


\begin{table*}
    \begin{threeparttable}
    \caption{The structure of four LORDs using the continuous dose space$^*$ (designs I, II, III, IV), and four LORDs using the discrete dose space$^{**}$ (designs I$^\prime$, II$^\prime$, III$^\prime$, IV$^\prime$), under four scenarios of  dose--response (A, B, C, D).} 
    \centering
    \footnotesize
    \begin{tabular}{ccccccccccccccccc}
    \hline
     & \multicolumn{16}{c}{Design} \\
    \cline{2-17}     
    Scenario  & \multicolumn{2}{c}{I} & \multicolumn{2}{c}{I$^\prime$} & 
    \multicolumn{2}{c}{II} & \multicolumn{2}{c}{II$^\prime$} & 
    \multicolumn{2}{c}{III} & \multicolumn{2}{c}{III$^\prime$} & 
    \multicolumn{2}{c}{IV} & \multicolumn{2}{c}{IV$^\prime$} \\
    \hline
    & $d_i$ & $\rho_i$ & $x_i$ & $\rho_i^\prime$ & $d_i$ & $\rho_i$ & $x_i$ & $\rho_i^\prime$ & $d_i$ & $\rho_i$ & $x_i$ & $\rho_i^\prime$ & $d_i$ & $\rho_i$ & $x_i$ & $\rho_i^\prime$\\
    \hline
      & -1.20 & 0.28 & $x_1$ & 0.28 & 0.92 & 0.45 & $x_3$ & 0.46 & -0.6 & 0.3 & $x_1$ & 0.11 & 0.92 & 0.55 & $x_3$ & 0.56 \\
    A & 2.32 & 0.36 & $x_4$ & 0.36 & 2.75 & 0.08 & $x_5$ & 0.08 & 3.86 & 0.7 & $x_2$ & 0.18 & 4.38 & 0.45 & $x_6$ & 0.11 \\
      & 4.38 & 0.36 & $x_7$ & 0.36 & 4.38 & 0.47 & $x_7$ & 0.46 & & & $x_6$ & 0.57 & & & $x_7$ & 0.33\\
      & & & & & & & & & & & $x_7$ & 0.14 & &  \\
    \hline
      &-1.20 & 0.25 & $x_1$ & 0.25 &-0.22 & 0.25 & $x_2$ & 0.25 &-1.20 & 0.09 & $x_1$ & 0.17 &-0.22 & 0.07 & $x_2$ & 0.05 \\
    B & -0.14 & 0.25 & $x_2$ & 0.25 & 0.54 & 0.25 & $x_3$ &    0.25 & 0.18 & 0.29 & $x_2$ & 0.25 & 0.76 & 0.46 &
      $x_3$ & 0.49 \\
      & 4.01 & 0.25 & $x_6$ & 0.25 & 4.01 & 0.25 & $x_6$ & 0.25 & 3.71 & 0.54 & $x_6$ & 0.51 & 3.71 & 0.41 & $x_6$ & 0.40 \\
      & 5.08 & 0.25 & $x_8$ & 0.25 & 5.08 & 0.25 & $x_8$ & 0.25 & 5.08 & 0.08 & $x_8$ & 0.07 & 5.08 & 0.06 & $x_8$ & 0.06 \\
    \hline
      & 2.08 & 0.25 & $x_4$ & 0.25 & 4.38 & 0.30 & $x_7$ & 0.29 & 5.07 & 0.61 & $x_8$ & 0.61 & 5.07 & 0.61 & $x_8$ & 0.61 \\
    C & 5.19 & 0.40 & $x_8$ & 0.40 & 5.27 & 0.28 & $x_8$ &    0.26 & 5.77 & 0.39 & $x_9$ & 0.39 & 5.77 & 0.39 &    
      $x_9$ & 0.39 \\
      & 5.77 & 0.35 & $x_9$ & 0.35 & 5.77 & 0.42 & $x_9$ & 0.45 & & & & & & & & \\
    \hline
      &-1.20 & 0.50 & $x_1$ & 0.50 & -1.02 & 0.50 & $x_1$ & 
      0.50 &-1.20 & 0.50 & $x_1$ & 0.50 &-1.02 & 0.50 & $x_1$ & 0.50 \\
    D &-0.11 & 0.50 & $x_2$ & 0.50 & 0.11 & 0.50 & $x_2$ &
      0.50 & 0.11 & 0.50 & $x_2$ & 0.50 & 0.11 & 0.50 & $x_2$ & 0.50 \\
    \hline
    \end{tabular}\label{Table2}
    \begin{tablenotes}
    \footnotesize
       \item[*] Continuous dose space: $\ln(\text{dose})\in\mathfrak{D}$ = [-1.20, 5.77] (on the original scale 0.3~mg to 320~mg).
       \item[**] $\mathfrak{X}$ contains nine doses: $x_1= $ -1.20, $x_2= $ -0.23, $x_3=0.92$, $x_4=2.02$, $x_5=3.00$, $x_6=3.69$, $x_7=4.38$, $x_8=5.08$, and $x_9=5.77$.  
    \end{tablenotes}    
    \end{threeparttable}
\end{table*}

\begin{table*}
    \begin{threeparttable}
    \caption{D-efficiency ($D_{\text{eff}}$) and c-efficiency ($c_{\text{eff}}$) of four LORDs using the continuous dose space$^*$ (designs I, II, III, IV), and four LORDs using the discrete dose space$^{**}$ (designs I$^\prime$, II$^\prime$, III$^\prime$, IV$^\prime$), under four scenarios of dose--response (A, B, C, D). For all designs, $D_{\text{eff}}$ is calculated relative to design~I, and $c_{\text{eff}}$ is calculated relative to design~III.}
    \centering
    \begin{tabular}{p{1.5cm}p{1.5cm}p{1.1cm}p{1.1cm}p{1.1cm}p{1.1cm}p{1.1cm}p{1.1cm}p{1.1cm}p{1.1cm}}
    \hline
    Scenario & \multicolumn{9}{c}{Design} \\
    \cline{3-10}     
      & & I & I$^\prime$ & II & II$^\prime$& III & III$^\prime$ & IV & IV$^\prime$ \\
    \cline{3-10} 
    A & $D_{\text{eff}}$ & 1 & 0.99 & 0.74 & 0.74 & 0.74 &
    0.77 & 0.70 & 0.72 \\
      & $c_{\text{eff}}$ & 0.9 & 0.88 & 0.78 & 0.79 & 1 &
      0.99 & 0.84 & 0.83 \\
    \hline
    B & $D_{\text{eff}}$ & 1.0 & 0.97 & 0.57 & 0.53 & 0.57 &
    0.77 & 0.36 & 0.33 \\
      & $c_{\text{eff}}$ & 0.66 & 0.63 & 0.42 & 0.38 & 1 & 
      0.89 & 0.63 & 0.62 \\
    \hline
    C & $D_{\text{eff}}$ & 1 & 0.99 & 0.6 & 0.58 & 0.43 & 
    0.42 & 0.43 & 0.42 \\
      & $c_{\text{eff}}$ & 0.72 & 0.74 & 0.72 & 0.73 & 1 & 
      1 & 1 & 1 \\
    \hline
    D & $D_{\text{eff}}$ & 1 & 0.7 & 0.9 & 0.7 & 1 & 0.7 &
    0.9 & 0.7 \\
      & $c_{\text{eff}}$ & 1 & 0.59 & 0.72 & 0.59 & 1 & 0.59 &
    0.72 & 0.59 \\
    \hline
    \end{tabular}\label{Table3}
    \begin{tablenotes}
    \footnotesize
       \item[*] Continuous dose space: $\ln(\text{dose})\in\mathfrak{D}$ = [-1.20, 5.77] (on the original scale 0.3~mg to 320~mg).
       \item[**] $\mathfrak{X}$ contains nine doses: $x_1= $ -1.20, $x_2= $ -0.23, $x_3=0.92$, $x_4=2.02$, $x_5=3.00$, $x_6=3.69$, $x_7=4.38$, $x_8=5.08$, and $x_9=5.77$.  
    \end{tablenotes}    
    \end{threeparttable}
\end{table*}

Figures~\ref{fig:scenarioA}, \ref{fig:scenarioB}, \ref{fig:scenarioC}, and \ref{fig:scenarioD} show the fulfillment of the GET conditions for the optimal designs in scenarios A, B, C, and D, respectively. 

We also observe from the Figures that the structure of optimal designs can vary depending on the true dose--response, the design space (interval or discrete set of doses), and the type of the chosen optimality criterion. To gain insights into the optimal designs, it is instructive to consider a few scenarios.

Consider Scenario~A in which $OBD$ falls in the middle part of the therapeutic window $[MinED, MTD]$. From Figure~\ref{fig:scenarioA} and Table~\ref{Table2}, Designs I and II are supported at three doses, one of which is $MTD$. 
Designs III and IV are supported at two doses, and both of them are different from $OBD$. 
The designs based on the discrete set of doses $\mathfrak{X}$ might differ in structure from those based on the continuum of doses $\mathfrak{D}$. For example, design III$^\prime$ is supported at four doses, $x_1=$ -1.20, $x_2=$ -0.23, $x_6=3.69$, and $x_7=4.38$, with allocation proportions of 0.11, 0.18, 0.57, and 0.17 at these doses. By contrast, design III is supported at two doses, $d_1=$ -0.60 and $d_2=3.86$, with allocation proportions of 0.3 and 0.7 at these doses. Overall, a given design on $\mathfrak{X}$ has similar values of $D_{\text{eff}}$ and $c_{\text{eff}}$ compared to the corresponding design on $\mathfrak{D}$ (Table~\ref{Table3}). If D- and c-efficiency are deemed equally important, we can use the Euclidean distance $\delta=\sqrt{(1-D_{\text{eff}})^2+(1-c_{\text{eff}})^2}$ as a measure of closeness of a design to the ``ideal'' point (1,1) in the D- and c-efficiency space, such that smaller values of $\delta$ indicate better tradeoff between D- and c-criteria. The values of $\delta$ for designs I, II, III, and IV are, respectively, 0.10, 0.34, 0.26, and 0.34, and the corresponding values for designs I$^\prime$, II$^\prime$, III$^\prime$, and IV$^\prime$ are, respectively, 0.12, 0.33, 0.23, and 0.33. This indicates that under Scenario~A, designs I and I$^\prime$ achieve the best balance between the goals of estimating the overall dose--response and $OBD$ while protecting patients from overly toxic doses. 

Consider Scenario~B with a very wide therapeutic window in which most of the doses have probability of success close to 1. From Figure~\ref{fig:scenarioB} and Table~\ref{Table2}, all optimal designs are supported at four doses, but their locations and the probability mass distributions are different. For designs I, I$^\prime$, II, and II$^\prime$, all optimal doses have probability mass of 0.25,  whereas for designs III, III$^\prime$, IV, and IV$^\prime$, the probability mass is mostly concentrated at the two middle doses. For designs II and IV, the sets of optimal doses include the lower and upper limits of the therapeutic window ($MinED$ and $MTD$), as well as two intermediate doses which are inflection points of the dose--success probability curve. The values of $\delta$ for designs I, II, III, and IV are, respectively, 0.34, 0.72, 0.43, and 0.74, and the corresponding values for designs I$^\prime$, II$^\prime$, III$^\prime$, and IV$^\prime$ are, respectively, 0.37, 0.78, 0.25, and 0.77. This indicates that under Scenario~B, design III$^\prime$ has, overall, best statistical properties.

Consider Scenario C in which most of the doses are safe, $MinED$ is in the upper part of the  dose range, $MTD$ is nearly at the end of the dose range, and $OBD$ is quite close to $MTD$. From Figure~\ref{fig:scenarioC} and Table~\ref{Table2}, designs I, I$^\prime$, II, and II$^\prime$ are supported at three doses, whereas designs III, III$^\prime$, IV, and IV$^\prime$ are supported at two doses. The values of $\delta$ for designs I, II, III, and IV are, respectively, 0.28, 0.49, 0.57, and 0.57, and the corresponding values for designs I$^\prime$, II$^\prime$, III$^\prime$, and IV$^\prime$ are, respectively, 0.26, 0.50, 0.58, and 0.58. This indicates that under Scenario~C, designs I and I$^\prime$ have, overall, best statistical properties.

Finally, consider Scenario D with a very narrow therapeutic window, with $\ln(MinED)=$ -1.02, $\ln(MTD)=0.11$, and $\ln(OBD)=$ -1.8, which is below the lower limit of the studied dose range $\mathfrak{D}=$ [-1.20, 5.77]. In this case, for the discrete set $\mathfrak{X}$ we have $\ln(MinED^\prime)=\ln(OBD^\prime)=x_1=$ -1.20, and $\ln(MTD^\prime)=x_2=$ -0.23. From Figure~\ref{fig:scenarioD} and Table~\ref{Table2}, one can see that all eight considered designs are supported at two doses (-1.20 and 0.11 for designs I and III; -1.02 and 0.11 for designs III and IV; and $x_1$ and $x_2$ for designs I$^\prime$, II$^\prime$, III$^\prime$, and IV$^\prime$). The values of $\delta$ for designs I, II, III, and IV are, respectively, 0, 0.3, 0, and 0.3, and the value of $\delta$ for designs I$^\prime$, II$^\prime$, III$^\prime$, and IV$^\prime$ is 0.51. Therefore, under Scenario~D, designs I and III are statistically most appealing if the investigator is willing to consider the continuum of doses; or, if the investigator wishes to use the discrete dose set, then an equal allocation at doses $x_1$ and $x_2$ provides the optimal strategy.

\section{Additional assessments of LORDs}\label{Sec4}
The above optimal designs were constructed under several assumptions and it is helpful to study their sensitivity to various misspecifications in the model assumptions before implementation.  Below, we consider three such situations.

\subsection{Sensitivity to the choice of the dose range}\label{Sec4.1}

Consider Scenario~C in which all doses are very safe and $MTD$ is located above the highest studied dose of the interval $\mathfrak{D}=$ [-1.20, 5.77] (Figure~\ref{fig:scenarioC}, top left plot).  Consider an extended  dose set with one extra dose, $x_{10}=6.91$, i.e.,  $\mathfrak{X}_e=\mathfrak{X}\cup\{x_{10}=6.91\}$ and the corresponding extended interval of doses $\mathfrak{D}_e=$ [-1.20, 6.91]. In this case, $\ln(MTD)=6.81$ with toxicity probability equal to $\Gamma=0.20$, whereas $\ln(MTD^\prime)=x_{10}=6.91$, because $x_{10}$ has toxicity probability 0.226, closest to the target value of $\Gamma=0.2$ among the doses from $\mathfrak{X}_e$. 

\begin{figure}
    \centering
    \includegraphics[width=0.95\linewidth]{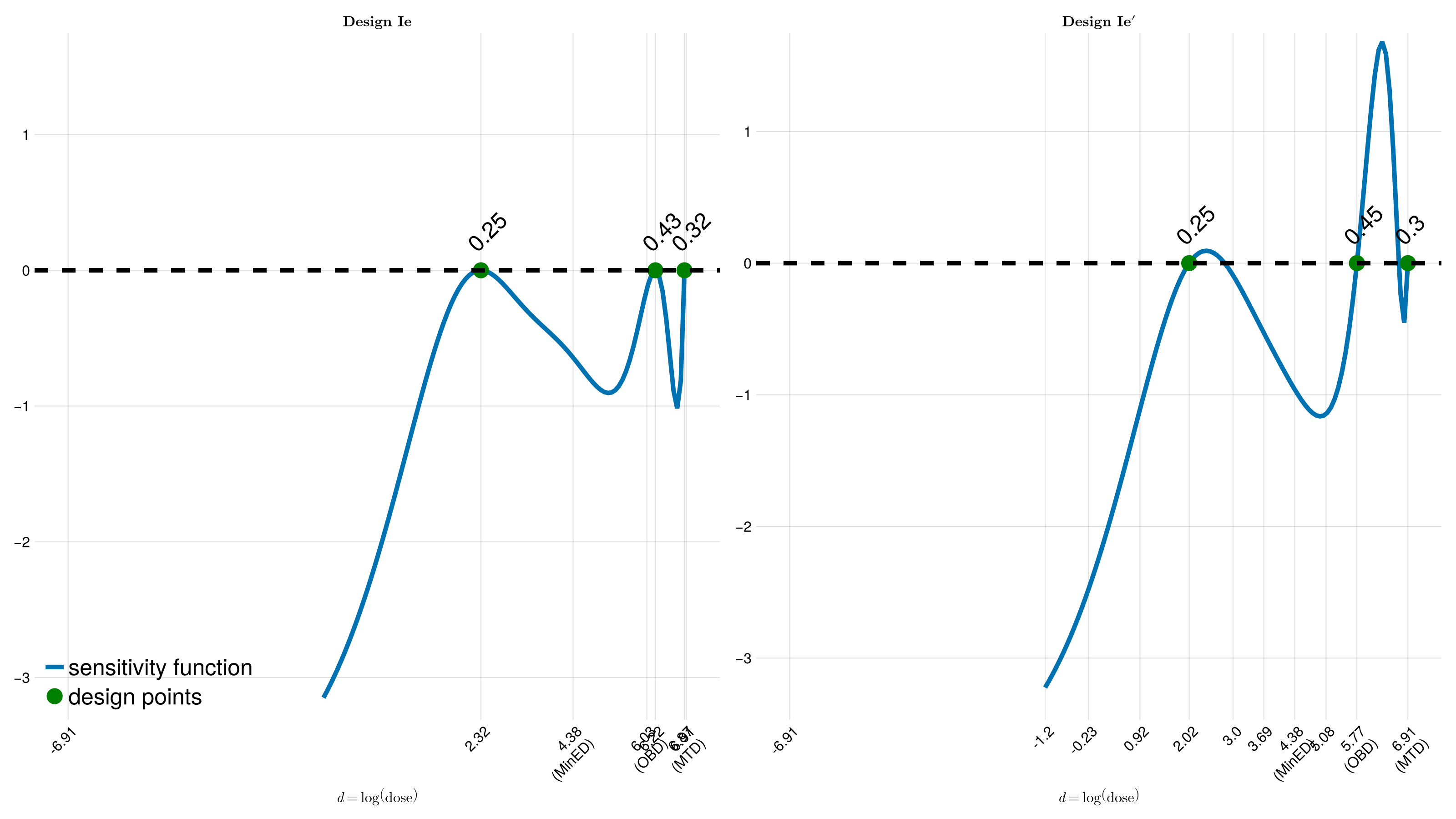}
    \caption{Optimal designs $\text{I}_e$ and $\text{I}_e^\prime$ for scenario C (cf. Section~\ref{Sec4.1}).}
    \label{fig:Sensitivity1}
\end{figure}

For this setup, we obtained design $\text{I}_e$ (D-optimal design under the restriction that the dose levels chosen from $\mathfrak{D}_e$ cannot exceed $\ln(MTD)=6.81$) and design $\text{I}^\prime_e$ (D-optimal design under the restriction that the dose levels chosen from $\mathfrak{X}_e$ cannot exceed $\ln(MTD^\prime)=x_{10}=6.91$). Table~\ref{Table4} shows the structure of designs $\text{I}_e$ and $\text{I}^\prime_e$ and their efficiencies relative to designs~I and I$^\prime$ (cf. Section~\ref{Sec3}), and Figure~\ref{fig:Sensitivity1} shows the fulfillment of the GET conditions for designs $\text{I}_e$ and $\text{I}^\prime_e$.

\begin{table*}
    \begin{threeparttable}
    \caption{The structure and efficiency of LORDs using extended dose spaces (design $\text{I}_e$ with $\mathfrak{D}_e$ = [-1.20, 6.91] and design  $\text{I}^\prime_e$ with $\mathfrak{X}_e=\mathfrak{X}\cup\{x_{10}=6.91\}$, in comparison with LORDs using the original dose spaces (design I with $\mathfrak{D}$ = [-1.20, 5.77]  and design  $\text{I}^\prime$ with $\mathfrak{X}=\{x_1,\ldots,x_9\}^*$), under Scenario~C.$^{**}$} 
    \centering
    \begin{tabular}{p{1.25cm}p{1.25cm}p{1.25cm}p{1.25cm}p{.5cm}p{1.9cm}p{1.25cm}p{1.9cm}p{1.25cm}}
    \hline
    \multicolumn{2}{c}{I} & \multicolumn{2}{c}{$\text{I}_e$} & & \multicolumn{2}{c}{$\text{I}^\prime$} & \multicolumn{2}{c}{$\text{I}^\prime_e$} \\
    \cline{1-4}\cline{6-9}     
     $d_i$ & $\rho_i$ & $d_i$ & $\rho_i$ & & $x_i$ & $\rho_i^\prime$ & 
     $x_i$ & $\rho_i^\prime$ \\
    \cline{1-4}\cline{6-9}  
      2.08 & 0.25 & 2.33 & 0.25 &  & $x_4=2.02$ & 0.25 & $x_4=2.02$ & 0.25 \\
      5.19 & 0.40 & 6.22 & 0.43 &  & $x_8=5.08$ & 0.40 & $x_9=5.77$ & 0.45 \\
      5.77 & 0.35 & 6.87 & 0.32 &  & $x_9=5.77$ & 0.35 & $x_{10}=6.91$ & 0.30 \\
    \hline
    \multicolumn{2}{c}{$D_{\text{eff}}$ ($\text{I}_e$ vs. I)} & \multicolumn{2}{c}{5.71} & & \multicolumn{2}{c}{$D_{\text{eff}}$ ($\text{I}_e^\prime$ vs. $\text{I}^\prime$)} & \multicolumn{2}{c}{5.36} \\
    \multicolumn{2}{c}{$c_{\text{eff}}$ ($\text{I}_e$ vs. I)} & \multicolumn{2}{c}{43.72} & & \multicolumn{2}{c}{$c_{\text{eff}}$ ($\text{I}_e^\prime$ vs. $\text{I}^\prime$)} & \multicolumn{2}{c}{27.86}\\     
    \hline
    \end{tabular}\label{Table4}
    \begin{tablenotes}
    \footnotesize
       \item[*] $\mathfrak{X}$ contains nine doses:  $x_1= $ -1.20, $x_2= $ -0.23, $x_3=0.92$, $x_4=2.02$, $x_5=3.00$, $x_6=3.69$, $x_7=4.38$, $x_8=5.08$, and $x_9=5.77$. 
       \item[**] For Scenario C, we have: 1) If the dose space is continuous ($\mathfrak{D}$ or $\mathfrak{D}_e$) $\Rightarrow$ $\ln(MinED)=4.38$, $\ln(OBD)=6.03$, and $\ln(MTD)=6.87$. 2) If the dose space is $\mathfrak{X}$ $\Rightarrow$ $\ln(MinED^\prime)=x_7$, and $\ln(OBD^\prime)=\ln(MTD^\prime)=x_9$. 3) If the dose space is $\mathfrak{X}_e=\mathfrak{X}\cup\{x_{10}=6.91\}$ $\Rightarrow$ $\ln(MinED^\prime_e)=x_7$, $\ln(OBD^\prime_e)=x_9$, and $\ln(OBD^\prime_e)=x_{10}$.
    \end{tablenotes}    
    \end{threeparttable}
\end{table*}

The three optimal doses for design $\text{I}_e$ are shifted to the right compared to those for design~I, while the dose allocation proportions for the two designs are similar. When comparing $\text{I}_e^\prime$ with  $\text{I}^\prime$, one can see that they have the same lowest dose ($x_4$) but the two other doses are shifted from $(x_8,x_9)$ for design~$\text{I}^\prime$ to $(x_9,x_{10})$ for design~$\text{I}_e^\prime$, with similar sets of allocation proportions for the two designs. There is a striking increase in estimation efficiency for the extended designs ($\sim$5.4-fold increase in D-efficiency and $\sim$28-fold increase in c-efficiency for $\text{I}_e^\prime$ compared to $\text{I}^\prime$).

\subsection{Sensitivity to the number and the values of the pre-specified dose levels}\label{Sec4.2}

Consider Scenario~A with a ``reduced'' set of five doses instead of nine doses, as considered originally. Define $\mathfrak{X}_r=\mathfrak{X}\setminus\{x_3,x_5,x_7,x_9\}=\{x_1,x_2,x_4,x_6,x_8\}$. For the original set $\mathfrak{X}$, $\ln(MTD^\prime)=x_7=4.38$ with the toxicity probability of 0.20, whereas for the reduced set $\mathfrak{X}_r$, $\ln(MTD^\prime_r)=x_6=3.69$ with the toxicity probability of 0.11, which is below the target level $\Gamma=0.20$. 

\begin{figure}
    \centering
    \includegraphics[width=0.95\linewidth]{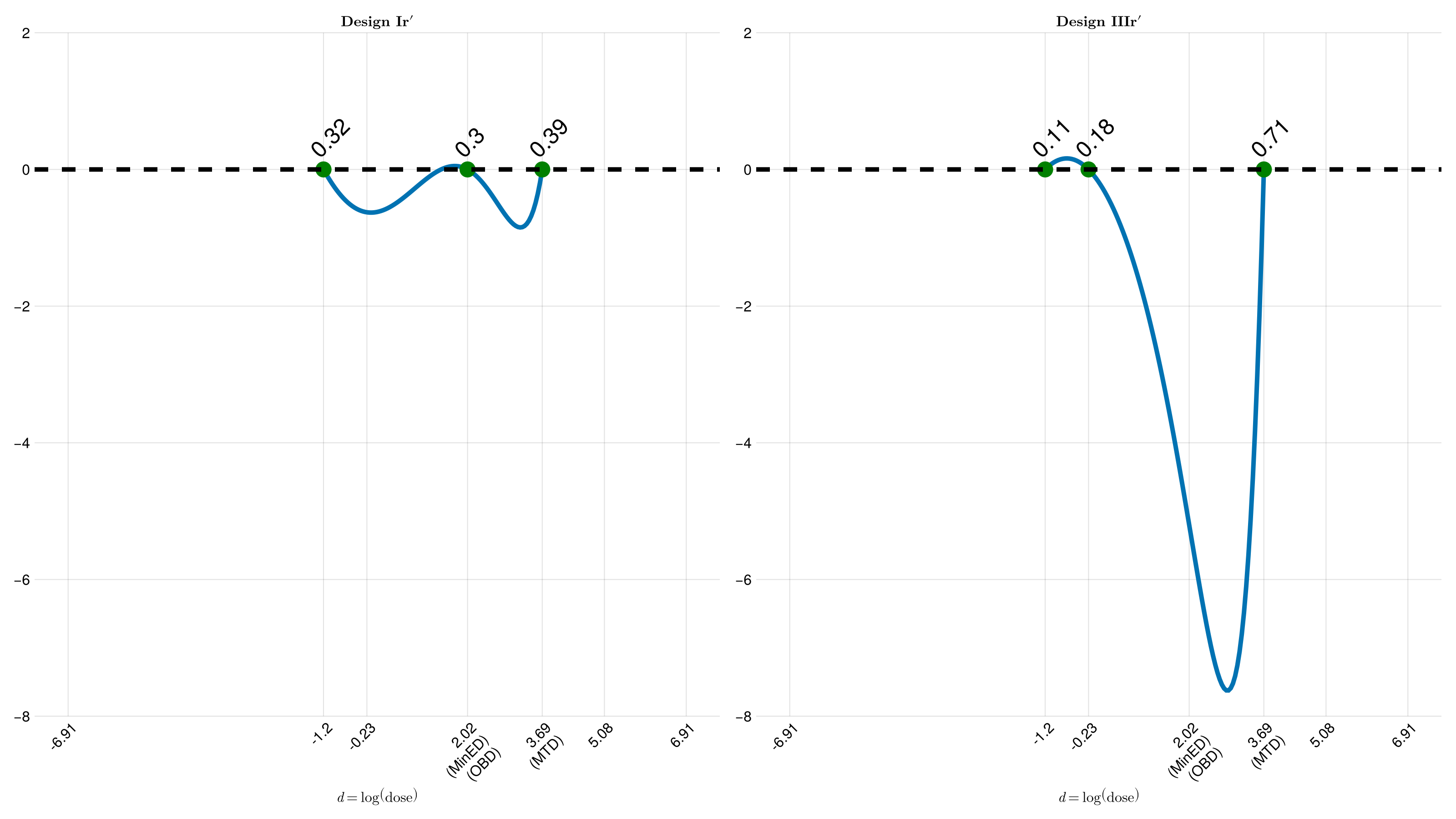}
    \caption{Optimal designs $\text{I}_r^\prime$ and $\text{III}_r^\prime$ for scenario A (cf. Section~\ref{Sec4.2}).}
    \label{fig:Sensitivity2}
\end{figure}

For this setup, we obtained design~$\text{I}^\prime_r$ (D-optimal design under the restriction that the dose levels chosen from $\mathfrak{X}_r$ cannot exceed $\ln(MTD^\prime)=x_6=3.69$) and design $\text{III}^\prime_r$ (c-optimal design under the restriction that the dose levels chosen from $\mathfrak{X}_r$ cannot exceed $\ln(MTD^\prime_r)=x_6=3.69$). Table~\ref{Table5} shows the structure of designs $\text{I}^\prime_r$ and $\text{III}^\prime_r$ and their efficiencies relative to designs~I$^\prime$ and III$^\prime$, respectively (cf.~Section~\ref{Sec3}), and Figure~\ref{fig:Sensitivity2} shows the fulfillment of the GET conditions for designs $\text{I}^\prime_r$ and $\text{III}^\prime_r$.

\begin{table*}
    \begin{threeparttable}
    \caption{The structure and efficiency of LORDs using the reduced dose space of five doses $\mathfrak{X}_r=\{x_1,x_2,x_4,x_6,x_8\}$ (designs $\text{I}_r^\prime$ and $\text{III}^\prime_r$), in comparison with LORDs using the original dose space of nine doses $\mathfrak{X}=\{x_1,\ldots,x_9\}^*$ (designs $\text{I}^\prime$ and $\text{III}^\prime$), under Scenario~A.$^{**}$} 
    \centering
    \begin{tabular}{p{1.9cm}p{1.25cm}p{1.9cm}p{1.25cm}p{.5cm}p{1.9cm}p{1.25cm}p{1.9cm}p{1.25cm}}
    \hline
    \multicolumn{2}{c}{$\text{I}^\prime$} & \multicolumn{2}{c}{$\text{I}_r^\prime$} & & \multicolumn{2}{c}{$\text{III}^\prime$} & \multicolumn{2}{c}{$\text{III}^\prime_r$} \\
    \cline{1-4}\cline{6-9}     
     $x_i$ & $\rho_i^\prime$ & $x_i$ & $\rho_i^\prime$ & & $x_i$ & $\rho_i^\prime$ & 
     $x_i$ & $\rho_i^\prime$ \\
    \cline{1-4}\cline{6-9}  
     $x_1=$ -1.20 & 0.28 & $x_1=$ -1.20 & 0.32 & & $x_1=$ -1.20 & 0.11 & $x_1=$ -1.20 & 0.10 \\
     $x_4=2.02$ & 0.36 & $x_4=2.02$ & 0.30 & & $x_2=$ -0.23 & 0.18 & $x_2=$ -0.23 & 0.19 \\
     $x_7=4.38$ & 0.36 & $x_6=3.69$ & 0.38 & & $x_6=3.69$ & 0.57 & $x_6=3.69$ & 0.71 \\
     & & & &  & $x_7=4.38$ & 0.14 & & \\
    \hline
    \multicolumn{2}{c}{$D_{\text{eff}}$ ($\text{I}_r^\prime$ vs. $\text{I}^\prime$)} & \multicolumn{2}{c}{0.78} & & \multicolumn{2}{c}{$D_{\text{eff}}$ ($\text{III}_r^\prime$ vs. $\text{III}^\prime$)} & \multicolumn{2}{c}{0.64} \\
    \multicolumn{2}{c}{$c_{\text{eff}}$ ($\text{I}_r^\prime$ vs. $\text{I}^\prime$)} & \multicolumn{2}{c}{0.61} & & \multicolumn{2}{c}{$c_{\text{eff}}$ ($\text{III}_r^\prime$ vs. $\text{III}^\prime$)} & \multicolumn{2}{c}{0.99}\\     
    \hline
    \end{tabular}\label{Table5}
    \begin{tablenotes}
    \footnotesize
       \item[*] $\mathfrak{X}$ contains nine doses:  $x_1= $ -1.20, $x_2= $ -0.23, $x_3=0.92$, $x_4=2.02$, $x_5=3.00$, $x_6=3.69$, $x_7=4.38$, $x_8=5.08$, and $x_9=5.77$. 
       \item[**] For Scenario A, we have: 1) If the dose space is $\mathfrak{X}$ $\Rightarrow$  $\ln(MinED^\prime)=x_3$, $\ln(OBD^\prime)=x_5$, and $\ln(MTD^\prime)=x_7$. 2) If the dose space is $\mathfrak{X}_r$ $\Rightarrow$ $\ln(MinED^\prime_r)=x_4$, $\ln(OBD^\prime_r)=x_4$, and $\ln(MTD^\prime_r)=x_6$. 
    \end{tablenotes}    
    \end{threeparttable}
\end{table*}

When comparing $\text{I}^\prime_r$ with $\text{I}^\prime$, one can see that both designs are supported at three doses, with two lowest doses in common ($x_1$ and $x_4$); however, $\text{I}^\prime_r$ has $x_6$ as the highest dose, whereas $\text{I}^\prime$ has $x_7$ as the highest dose. When comparing $\text{III}^\prime_r$ with $\text{III}^\prime$, one can see that these two designs have three doses in common ($x_1$, $x_2$, and $x_6$); however, $\text{III}^\prime$ has an extra dose, $x_7$ (with a 14\% allocation proportion), which is not available in $\mathfrak{X}_r$. As one can see from Table~\ref{Table5}, there is some loss in both D- and c-efficiency for LORDs using $\mathfrak{X}_r$ compared to the corresponding LORDS using $\mathfrak{X}$.

\subsection{Sensitivity to the definition of the target doses}\label{Sec4.3}

Suppose the investigator wishes to define $MinED$ differently from (\ref{MinED}). Instead of expressing it in terms of the probability of a neutral outcome, they define it in terms of the conditional probability of efficacy given no toxicity, that is:
\begin{equation}\label{MinED.star}
    MinED_*=\underset{d}{\arg}\{p_{E|\bar{T}}(d,\thetaB)=0.6\},
\end{equation}where $p_{E|\bar{T}}(d,\thetaB)$ is given by (\ref{logistic1}). Likewise, define 
\begin{equation}\label{MinED.star.prime}
    MinED_*^\prime=\arg\underset{1\le i\le K}{\min}|p_{E|\bar{T}}(x_i,\thetaB)-0.6|,
\end{equation}where $x_i\in\mathfrak{X}$, $i=1,\ldots,K$. 

With such definitions, for Scenario~A, we have $\ln(MinED_*)=$ -0.79, which is smaller than the originally considered $\ln(MinED)=0.92$. The newly defined therapeutic window $[\ln(MinED_*),\ln(MTD)]=$ [-0.79, 4.38] is wider than the originally considered $TW=[\ln(MinED),\ln(MTD)]=[0.92,4.38]$. Also, in this case, $\ln(MinED_*^\prime)=x_1=$ -1.2 with $p_{E|\bar{T}}(x_1,\thetaB)\approx0.54$ (note that for $x_2=$ -0.23, $p_{E|\bar{T}}(x_2,\thetaB)\approx0.67$, which is ``more distant'' from 0.6 than 0.54, the value of $p_{E|\bar{T}}$ at dose $x_1$), and $\ln(MTD^\prime)=x_7=4.38$.

\begin{figure}
    \centering
    \includegraphics[width=0.95\linewidth]{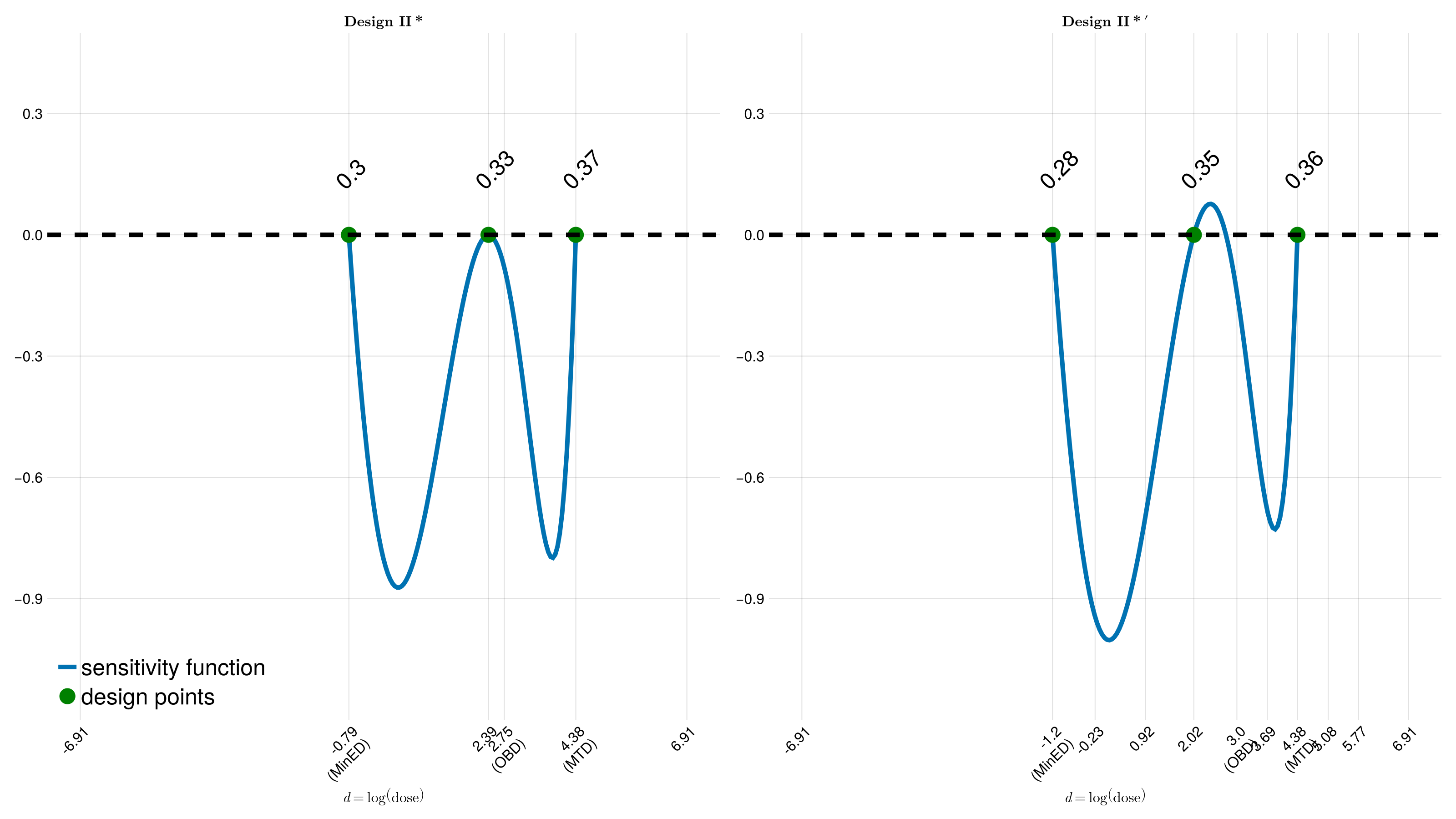}
    \caption{Optimal designs $\text{II}_*$ and $\text{II}_*^\prime$ for scenario A (cf. Section~\ref{Sec4.3}).}
    \label{fig:Sensitivity3}
\end{figure}

For the considered setup, we obtained design $\text{II}_*$ (D-optimal design under the restriction that the dose levels from the continuum of doses should be within $[\ln(MinED_*),\ln(MTD)]$), and design $\text{II}_*^\prime$ (D-optimal design under the restriction that the dose levels from the discrete set $\mathfrak{X}=\{x_1,\ldots,x_9\}$ should be within $[\ln(MinED_*^\prime),\ln(MTD^\prime)]$). Table~\ref{Table6} shows the structure of designs $\text{II}_*$ and $\text{II}_*^\prime$ and their efficiencies relative to designs~II and II$^\prime$, respectively (cf.~Section~\ref{Sec3}), and Figure~\ref{fig:Sensitivity3} shows the fulfillment of the GET conditions for $\text{II}_*$ and $\text{II}_*^\prime$.

\begin{table*}
    \begin{threeparttable}
    \caption{The structure and efficiency of LORDs using $\mathfrak{D}=$ [-1.20, 5.77] with $MinED$ defined by (\ref{MinED}) (design II) and $MinED_*$ defined by (\ref{MinED.star}) (design II$_*$), and LORDs using $\mathfrak{X}=\{x_1,\ldots,x_9\}^*$ with $MinED^\prime$ defined by (\ref{MinED.prime}) (design II$^\prime$) and $MinED_*^\prime$ defined by (\ref{MinED.star.prime}) (design II$_*^\prime$), under Scenario A.$^{**}$} 
    \centering
    \begin{tabular}{p{1.9cm}p{1.25cm}p{1.9cm}p{1.25cm}p{.5cm}p{1.9cm}p{1.25cm}p{1.9cm}p{1.25cm}}
    \hline
    \multicolumn{2}{c}{II} & \multicolumn{2}{c}{$\text{II}_*$} & & \multicolumn{2}{c}{$\text{II}^\prime$} & \multicolumn{2}{c}{$\text{II}^\prime_*$} \\
    \cline{1-4}\cline{6-9}     
     $d_i$ & $\rho_i$ & $d_i$ & $\rho_i$ & & $x_i$ & $\rho_i^\prime$ & 
     $x_i$ & $\rho_i^\prime$ \\
    \cline{1-4}\cline{6-9}  
     0.92 & 0.45 & -0.79 & 0.30 & & $x_3=0.92$ & 0.46 & $x_1=$ -1.20 & 0.28 \\
     2.75 & 0.09 &  2.39 & 0.33 & & $x_5=3.00$ & 0.08 & $x_4=2.02$   & 0.36 \\
     4.38 & 0.47 &  4.38 & 0.37 & & $x_7=4.38$ & 0.46 & $x_7=4.38$   & 0.36 \\
    \hline
    \multicolumn{2}{c}{$D_{\text{eff}}$ ($\text{II}_*$ vs. II)} & \multicolumn{2}{c}{1} & & \multicolumn{2}{c}{$D_{\text{eff}}$ ($\text{II}_*^\prime$ vs. II$^\prime$)} & \multicolumn{2}{c}{1} \\
    \multicolumn{2}{c}{$c_{\text{eff}}$ ($\text{II}_*$ vs. II)} & \multicolumn{2}{c}{1} & & \multicolumn{2}{c}{$c_{\text{eff}}$ ($\text{II}_*^\prime$ vs. II$^\prime$)} & \multicolumn{2}{c}{1}\\     
    \hline
    \end{tabular}\label{Table6}
    \begin{tablenotes}
    \footnotesize
       \item[*] $\mathfrak{X}$ contains nine doses:  $x_1= $ -1.20, $x_2= $ -0.23, $x_3=0.92$, $x_4=2.02$, $x_5=3.00$, $x_6=3.69$, $x_7=4.38$, $x_8=5.08$, and $x_9=5.77$. 
       \item[**] For Scenario A, we have: 1) If the dose space is $\mathfrak{D}=$ [-1.20, 5.77] $\Rightarrow$  
       $\ln(MinED)=\underset{d}{\arg}\{\pi_0(d,\thetaB)=0.2\}=0.92$, and $\ln(MinED_*)=\underset{d}{\arg}\{p_{E|\bar{T}}(d,\thetaB)=0.6\}=$ -0.79. Also, in this case, $\ln(OBD)=2.75$ and $\ln(MTD)=4.38$. 2) If the dose space is $\mathfrak{X}=\{x_1,\ldots,x_9\}$ $\Rightarrow$  
       $MinED^\prime=\underset{1\le i\le 9}{\arg\min}|\pi_0(x_i,\thetaB)-0.2|=x_3$, and $MinED^\prime_*=\underset{1\le i\le 9}{\arg\min}{|p_{E|\bar{T}}(x_i,\thetaB)-0.6|}=x_1$. Also, in this case, $\ln(OBD^\prime)=x_5$ and $\ln(MTD^\prime)=x_7$.
    \end{tablenotes}    
    \end{threeparttable}
\end{table*}


When comparing $\text{II}_*$ with II, one can see that both designs are supported at three doses: 1) the minimum efficacious dose (which is different for $\text{II}_*$ and II); 2) the intermediate dose (which coincides with $\ln(OBD)=2.75$ for design II, or it is 2.39 for design $\text{II}_*$); and 3) the maximum tolerated dose ($\ln(MTD)=4.38$ for both designs). When comparing $\text{II}_*^\prime$ with $\text{II}^\prime$, one can see that both designs are also supported at three doses, with the highest dose in common ($x_7=4.38$), but two other doses shifted from $(x_3,x_5)$ for design~$\text{II}^\prime$ to $(x_1,x_4)$ for design~$\text{II}_*^\prime$. Both D- and c-efficiency are identical for the pairs of designs (II and $\text{II}_*$) and ($\text{II}^\prime$ and $\text{II}_*^\prime$).

\subsection{Comparing LORDs with a more heuristic design}\label{Sec4.4}

In this section, we compare the performance of LORDs with another, more heuristically chosen design. Various phase~I/II trial designs are available \citep{Yuan2016book, Zhou2019}. To fix ideas, we consider only one design, the random walk rule (RWR) to target safe $OBD$ \citep{Ivanova2003} because this design is very simple and has established statistical properties. For RWR, let $(\delta_j,\UB_j)$ denote the dose level and the outcome of the $j$th participant in the trial, $j\ge 1$. Suppose the assigned dose is $\delta_j=x_m\in\mathfrak{X}$. The \emph{dose recommendation} for the $(j+1)$st patient, $r_{j+1}$, is as follows:
\begin{equation*}
    r_{j+1}=
    \begin{cases} 
        x_{m+1},\text{ if  } \UB_j=(1,0,0)^\top;\\ 
        x_{m},\text{ if  } \UB_j=(0,1,0)^\top;\\
        x_{m-1}\text{ if  } \UB_j=(0,0,1)^\top.
    \end{cases}
\end{equation*}In addition, let $\hat{d}_\Gamma\in\mathfrak{X}$ be the maximum dose level at which the estimated probability of toxicity based on data from $j$ patients is less than $\Gamma$. Then the \emph{dose assignment} for the the $(j+1)$st patient is made as $\delta_{j+1}=\min(r_{j+1},\hat{d}_\Gamma)$. Appropriate modifications are made at the lowest and the highest doses.  

\begin{figure}
    \centering
    \includegraphics[width=0.95\linewidth]{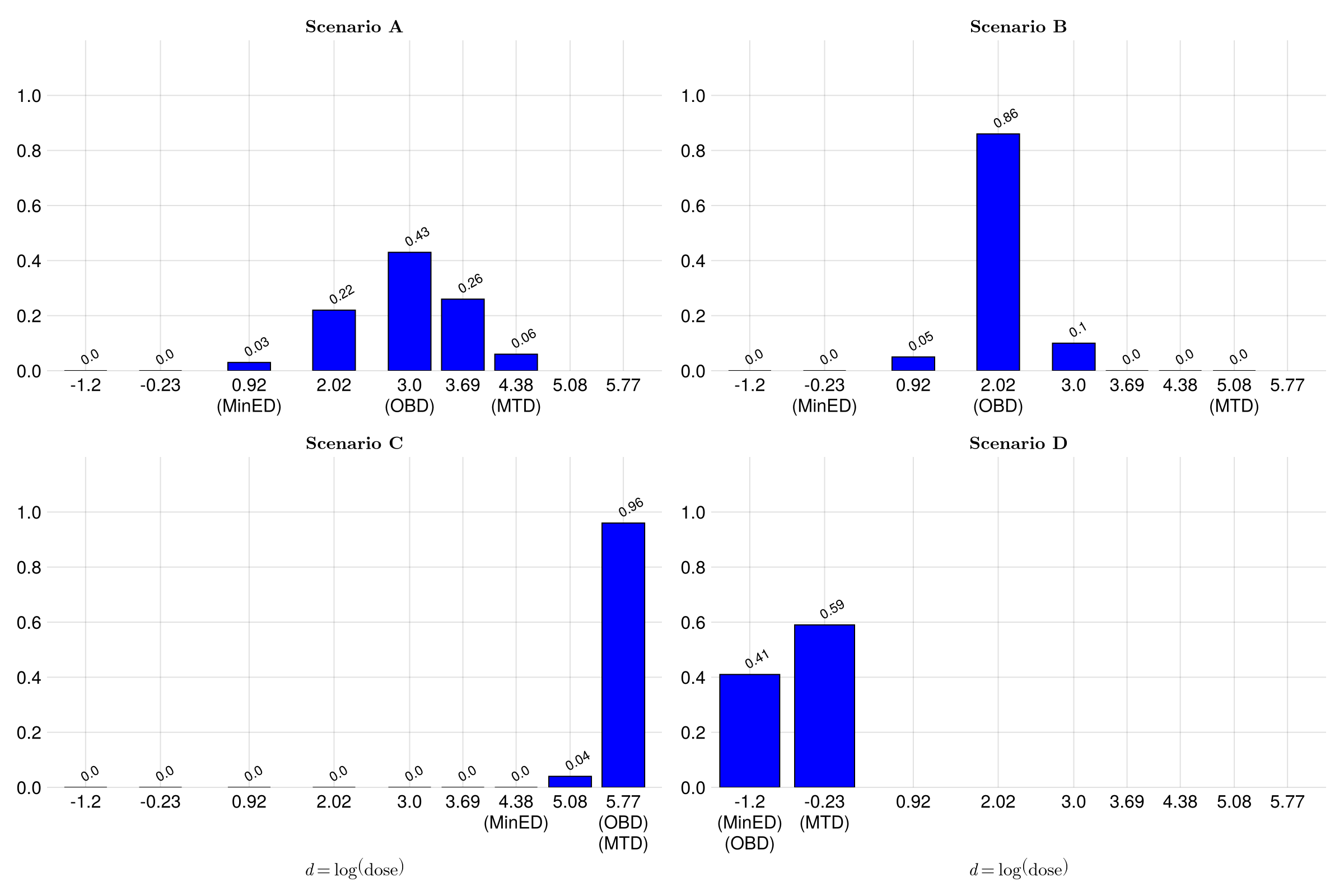}
    \caption{Stationary distribution of RWR \citep{Ivanova2003} for Scenarios A, B, C, and D (cf. Section~\ref{Sec4.4}).}
    \label{fig:RWR}
\end{figure}

RWR induces a Markov chain on the lattice of doses in $\mathfrak{X}$, with a stationary distribution that can be calculated for a given set of toxicity--efficacy probabilities \citep{Ivanova2003}. Let $x_\ell$ denote the largest dose in $\mathfrak{X}$ with $\pi_2(x_\ell,\thetaB)\le\Gamma$. Then the stationary distribution of the RWR design targeting safe $OBD$ has the following allocation probabilities ($\Pi_i$) at doses $x_i$, $i=1,\ldots,\ell$:
\begin{equation}\label{stationary}
\Pi_i=\prod_{j=1}^i\lambda_j,\quad\lambda_1=\left(1+\sum_{k=2}^\ell\prod_{j=2}^k\lambda_j\right)^{-1},\quad\lambda_k=\frac{\pi_0(x_{k-1},\thetaB)}{\pi_2(x_k,\thetaB)},
\end{equation}where $k=2,\ldots,\ell$. Figure~\ref{fig:RWR} shows this distribution under Scenarios A, B, C, and D. 

Let $\xiB_{\text{RWR}}=\{(x_i,\Pi_i),\;i=1,\ldots,K\}$, where $x_i\in\mathfrak{X}$, $\Pi_i\in[0,1]$, and $\sum_{i=1}^K\Pi_i=1$. Then its FIM is $\MB(\xiB_{\text{RWR}},\thetaB)=\sum_{i=1}^K\Pi_i\muB(x_i,\thetaB)$, and one can calculate the D-efficiency and c-efficiency of $\xiB_{\text{RWR}}$ relative to other designs. For a proper comparison, we consider RWR along with LORDs I$^\prime$, II$^\prime$, III$^\prime$, and IV$^\prime$ on the discrete set $\mathfrak{X}=\{x_1,\ldots,x_9\}$, as described in Section~\ref{Sec3}. Under a given scenario for dose--response, we consider three measures of ``worth'' of a design---D-efficiency ($D_{\text{eff}}$), c-efficiency ($c_{\text{eff}}$), and the expected proportion of successes in the trial. For a given design, $D_{\text{eff}}$ is calculated relative to ``benchmark'' design I$^\prime$, and $c_{\text{eff}}$ is calculated relative to ``benchmark'' design III$^\prime$; see (\ref{D_efficiency}) and (\ref{c_efficiency}).

The ``ethical worth'' of a design $\xiB$ is calculated as the expected proportion of successes relative to the ``most ethical,'' unrealistic design that assigns all patients to the true $OBD^\prime$ \citep{HardwickStout2003}:
\begin{equation}
    s = \left\{\sum_{i=1}^K\xi_i\pi_1(x_i,\thetaB)\right\}/\pi_1(OBD^\prime,\thetaB),
\end{equation}where $\xi_i$ is the allocation proportion for the $i$th dose ($\xi_i=\Pi_i$ for RWR and $\xi_i=\rho_i^\prime$ for designs I$^\prime$, II$^\prime$, III$^\prime$, and IV$^\prime$). 

\begin{figure}
    \centering
    \includegraphics[width=1\linewidth]{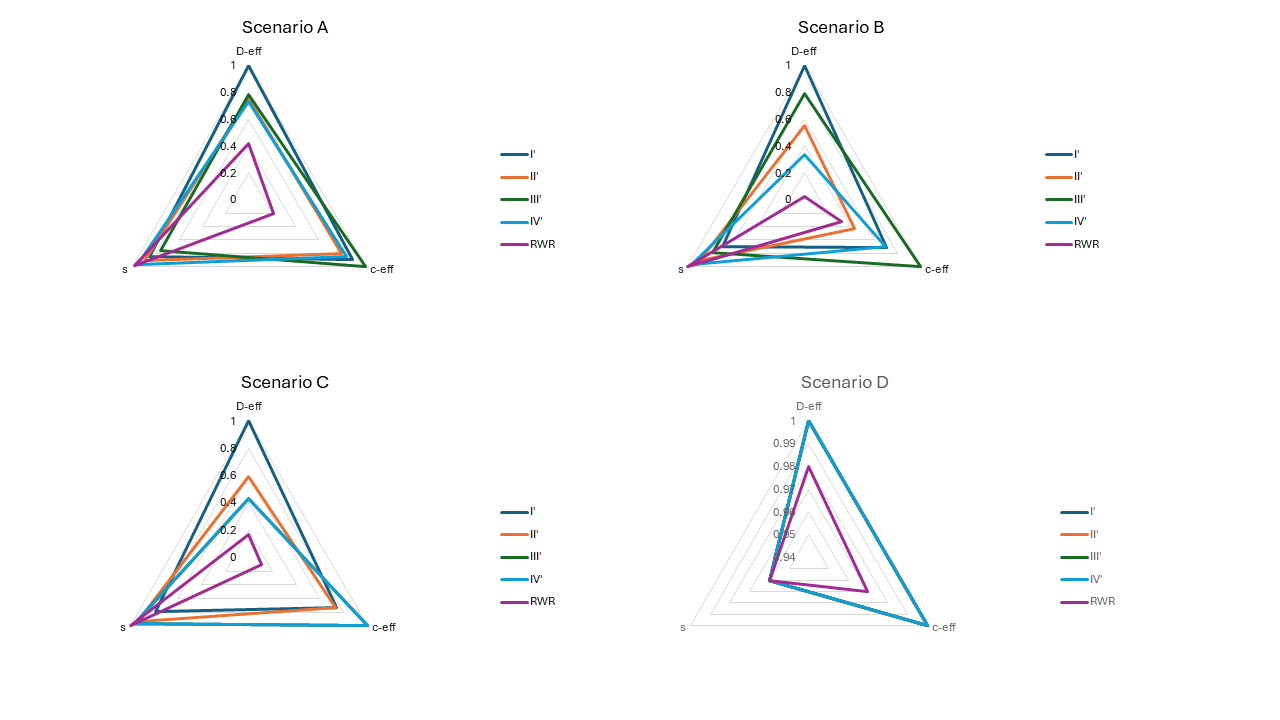}
    \caption{Radar plots for comparing designs $\text{I}^\prime$, $\text{II}^\prime$, $\text{III}^\prime$, $\text{IV}^\prime$, and RWR, with respect to D-efficiency ($D_{\text{eff}}$), c-efficiency ($c_{\text{eff}}$), and the expected proportion of successes relative to the ``most ethical'' design that assigns 100\% of patients to $OBD^\prime$ ($s$), under Scenarios A, B, C, and D (cf. Section~\ref{Sec4.4}).}
    \label{fig:DesignComparison}
\end{figure}

Table~\ref{Table7}  provides a comparative summary of five designs, and Figure~\ref{fig:DesignComparison} displays the radar plots of the five designs under the four scenarios of dose--response. The radar plots visualize the tradeoff among the competing criteria of D-efficiency for estimating the overall dose--response, c-efficiency for estimating $OBD^\prime$, and the ``ethical'' worth of a design relative to the benchmark design that treats all study participants with the $OBD^\prime$. 

\begin{table*}
    \begin{threeparttable}
    \caption{A comparative summary of designs $\text{I}^\prime$, $\text{II}^\prime$, $\text{III}^\prime$, $\text{IV}^\prime$, and RWR using dose set $\mathfrak{X}^*$ under four scenarios of dose-response (A, B, C, D). For all designs, $D_{\text{eff}}$ is calculated relative to design~$\text{I}^\prime$, $c_{\text{eff}}$ is calculated relative to design~$\text{III}^\prime$, and $s$ is calculated as the expected proportion of successes in the trial relative to the ``idealized'' design that assigns all patients to $OBD^\prime$. Score$^{**}$ is a measure of the overall performance of the design (larger is better).} 
    \centering
    \begin{tabular}{p{1.5cm}p{1.5cm}p{2cm}p{2cm}p{2cm}p{2cm}p{2cm}}
    \hline
    Scenario & \multicolumn{6}{c}{Design} \\
    \cline{3-7}     
      & & I$^\prime$ & II$^\prime$ & III$^\prime$ & IV$^\prime$ & RWR \\
    \cline{3-7} 
    A  & $D_{\text{eff}}$ & 1 & 0.75 & 0.78 & 0.73 & 0.42 \\
       & $c_{\text{eff}}$ & 0.89 & 0.80 & 1 & 0.84 & 0.21 \\
       & $s$ & 0.85 & 0.90 & 0.76 & 0.96 & 0.98 \\
       & Score  & 0.91 & 0.82 & 0.84 & 0.84 & 0.48 \\
    \hline
    B  & $D_{\text{eff}}$ & 1 & 0.55 & 0.79 & 0.34 & 0.02 \\
       & $c_{\text{eff}}$ & 0.71 & 0.43 & 1 & 0.70 & 0.32 \\
       & $s$ & 0.70  & 0.90 & 0.79 & 0.97 & 1 \\
       & Score  & 0.80 & 0.61 & 0.86 & 0.64 & 0.34 \\
    \hline   
    C  & $D_{\text{eff}}$ & 1 & 0.59 & 0.43 & 0.43 & 0.17 \\
       & $c_{\text{eff}}$ & 0.74 & 0.73 & 1 & 1 & 0.11 \\
       & $s$ & 0.79 & 0.94 & 0.97 & 0.97 & 1 \\
       & Score   & 0.84 & 0.75 & 0.78 & 0.78 & 0.32 \\
    \hline
    D  & $D_{\text{eff}}$ & 1 & 1 & 1 & 1 & 0.98 \\
       & $c_{\text{eff}}$ & 1 & 1 & 1 & 1 & 0.97 \\
       & $s$ & 0.96 & 0.96 & 0.96 & 0.96 & 0.96 \\
       & Score   & 0.99 & 0.99 & 0.99 & 0.99 & 0.97 \\
    \hline
    \end{tabular}\label{Table7}
    \begin{tablenotes}
    \footnotesize
       \item[*] $\mathfrak{X}$ contains nine doses:  $x_1= $ -1.20, $x_2= $ -0.23, $x_3=0.92$, $x_4=2.02$, $x_5=3.00$, $x_6=3.69$, $x_7=4.38$, $x_8=5.08$, and $x_9=5.77$.  
       \item[**] Score = $\{(D_{\text{eff}}\cdot c_{\text{eff}} + D_{\text{eff}}\cdot s + c_{\text{eff}}\cdot s)/3\}^{1/2}$. It measures the average impact of all pairs of factors on the overall performance of the design.   
    \end{tablenotes}    
    \end{threeparttable}
\end{table*}

One observes that LORDs have, overall, superior performance compared to RWR across all considered scenarios. While RWR prioritizes the treatment goal and concentrates dose assignments at and around $OBD^\prime$, it gas poor D- and c-efficiency in all scenarios, except for Scenario~D, where RWR is very similar to LORDs. Among the considered LORDs, design~I$^\prime$ has the best performance, followed by design~III$^\prime$.

\section{Discussion and future work}\label{Sec5}

In the current work, we have explored the utility of various Locally Optimal Restricted Designs (LORDs) for phase~I/II dose-finding studies. LORDs require a statistical model for dose--toxicity--efficacy relationship, and we used a four-parameter CR model \citep{FanChaloner2001, FanChaloner2004} for this purpose. We considered several important dimensions of the optimal design problem---the choice of a design space (an interval or a discrete set of doses), two statistical criteria (D-optimality and c-optimality), and two different sets of ``ethical'' restrictions on the study doses (using $MTD$ as the upper bound on the dose range or using the therapeutic window [$MinED, MTD$] as the desirable dose range). With these considerations, we formulated and solved various optimal design problems. We used PSO \citep{pso1995, Luke2013Metaheuristics} to find the optimal designs and the General Equivalence Theorem \citep{KieferWolfowitz1960} to verify the optimality of the obtained solutions. We investigated four scenarios of dose--response, informed by \cite{Whitehead2004} to illustrate the structure and efficiency of the LORDs. We also investigated the sensitivity of the results to different modifications of the design problem, such as extending or reducing the dose space or using a different definition for one of the target doses ($MinED$). Furthermore, we illustrated by example how LORDs can be used at the study planning stage to facilitate a comparison with some other competitive design options.

Our key finding is that LORDs provide statistically efficient and ethical benchmark designs with robust performance across the considered experimental scenarios of CR dose--response. Varying some aspects of the design has an impact on the structure and efficiency of LORDs. When evaluated simultaneously with respect to D-efficiency, c-efficiency, and the expected proportion of successes, LORDs exhibited superior performance compared to the random walk rule (RWR) targeting safe $OBD$ \citep{Ivanova2003}. While our comparison included only one design (RWR), it can be easily done for other phase~I/II designs such as BOIN12 \citep{Lin2020}, U-BOIN \citep{Zhou2019}, etc. It is important to note that our comparison should be interpreted in the ``idealized'' context, assuming that the CR model is plausible, the true values of the model parameters are known, and the asymptotic (large sample) assumptions apply. 

LORDs have several limitations. First, they depend on a statistical model, which may be misspecified in a number of ways. Instead of a CR for the trinomial outcome, one can consider other contingent response models \citep{RabieFlournoy2013}, e.g., the positive-negative extreme value model, and obtain similar results. If there are several candidate models for the dose--response relationship, one can consider $T$-optimal designs for model discrimination \citep{Dette2009} to mitigate the problem of misspecification of the distributional assumptions. Importantly, PSO can be applied to any considered statistical model, and the objective function does not need to be expressed in a closed form.  

Second, even if the considered model is plausible, LORDs depend on the true values of the model parameters that are unknown at the outset. To address this limitation, one can consider an adaptive design, e.g., performing initial dose assignments using some ``start-up'' design such as RWR to ascertain data for estimating the model parameters, and performing subsequent dose assignments according to the estimated optimal design \citep{DragalinFedorov2006}. While such an approach is scientifically sound, it requires careful calibration of the design parameters, including the choice of a ``start-up'' design, the choice of a sample size for different stages of the experiment, the rules for handling situations when the model parameters cannot be estimated for a given dataset at an interim analysis, the study stopping rules, etc. We regard the construction of adaptive versions of LORDs as an important future work.

Another important issue we have not addressed in the current paper is the ``optimal'' choice of the set of dose levels $\mathfrak{X}=\{x_1,\ldots,x_K\}$. We assumed that $\mathfrak{X}$ is provided by the investigator. In practice, investigators select dose levels based on previous experiments and manufacturing constraints. If the number of doses in $\mathfrak{X}$ is sufficiently large and they are located close to the unknown upfront target doses, then LORDs can be highly efficient. On the other hand, if the lattice of doses is sparse and/or these doses are ``far away'' from the target doses, then design efficiency may be lost, and there is a risk of selecting ``wrong'' doses at the end of the study. How to formally justify the choice of $\mathfrak{X}$ for a given experiment merits further investigation. A useful starting point is to identify LORDs based on the interval of doses $\mathfrak{D}$ and compare them with LORDs derived from the discrete set $\mathfrak{X}\subset\mathfrak{D}$. If the efficiency of the designs based on $\mathfrak{X}$ is high, no changes to $\mathfrak{X}$ are necessary. However, if the efficiency is low, the dose levels in $\mathfrak{X}$ may need further refinement and calibration, assuming the manufacturing of these new doses is feasible.

It should be noted that LORDs are different from penalized optimal designs (PODs) \citep{DragalinFedorov2006} which maximize the chosen statistical criterion while penalizing doses that are too toxic and/or have poor efficacy. The utility of PODs was investigated by \cite{DragalinFedorov2006, Dragalin2008, Pronzato2010, GaoRosenberger2013, Alam2019}, among others, who theoretically and through simulations affirmed the competitive performance of (adaptive) PODs compared to some more heuristically chosen designs. Unlike PODs, LORDs impose restrictions directly on the dose range, for example, by requiring that the study doses cannot exceed the unknown MTD, which depends on the model parameters. In this sense, LORDs are conceptually similar to the designs considered by \cite{Mats1998} and \cite{Haines2003} in the context of phase~I dose-toxicity studies.

In conclusion, LORDs provide useful benchmarks for statistical estimation efficiency under some ethical restrictions common to dose-finding studies. More research is needed to better understand the utility of LORDs and their extensions for phase~I/II dose-finding experiments.

\section*{Acknowledgement} The research of Wong is partially supported by  the Yushan Fellow Program by the Ministry of Education (MOE), Taiwan. (MOE-108-YSFMS-0004-012-P1).

\section*{Appendix: Particle Swarm Optimization algorithm}

For a given $OFV=f(P)$, the algorithm works as follows:

\begin{enumerate}
    \item initialize a \textit{swarm size}, $S$, i.e., a number of particles in a swarm.
    \item initialize \textit{maximum number of iterations}, $N_{\max}$.
    \item set an \textit{inertia weight} value, $w$. This PSO parameter can be fixed or may change over the iterations.
    \item set a \textit{cognitive coefficient} value, $c_1$.
    \item set a \textit{social coefficient} value, $c_2$.
    \item initialize a \textit{swarm} of particles (\textit{particles' positions}) at random:
    $$
    P^{(i)}=\left(p_1, \ldots, p_n\right), \: i = 1, \ldots, S.
    $$
    \item for each particle in the swarm, initialize velocities \textit{at random}: 
    $$
    V^{(i)} = (v_1, \ldots, v_n), \: i = 1, \ldots, S.
    $$
    \item for each particle, evaluate 
    $OVF^{(i)} = f(P^{(i)})$; let 
    $$
    \begin{aligned}
    OVF^{(i)}_{best} &= OVF^{(i)} \\
    P^{(i)}_{best} &= P^{(i)}
    \end{aligned}, \: i = 1, \ldots, S.
    $$
    \item evaluate swarm's the ``best'' OFV and position:
    $$
    \begin{aligned}
    OFV_{global} &= \max_{i = 1, \ldots, S}{OVF^{(i)}}\\
    P_{global} &= P^{(i^*)}, \text{ where }i^*:\: OVF^{(i^*)} = OFV_{global}. 
    \end{aligned}
    $$
    \item at each iteration step
    \begin{itemize}
        \item \textit{move the swarm} towards the ``\textit{best solution}'':
        \begin{itemize}
            \item for each particle, generate two random vectors, $\mathbf{r}^{(i)}_1$ and $\mathbf{r}^{(i)}_2$ of the same size as particles in the swarm.
            \item update particles' velocities:
            $$
            V^{(i)} = wV^{(i)} + c_1\mathbf{r}^{(i)}_1\left(P^{(i)}_{best} - P^{(i)}\right) + c_2\mathbf{r}^{(i)}_2\left(P_{global} - P^{(i)}\right), \:  i =1, \ldots, S.
            $$
            \item update particles' positions:
            $$
            P^{(i)} = P^{(i)} + V^{(i)}, \:  i =1, \ldots, S.
            $$
        \end{itemize}
        \item update an \textit{individual best} position of each particle, $P^{(i)}_{best}$ ($i = 1, \ldots, S$), and a \textit{global best} position, $P_{global}$;
        \item update an \textit{individual best} OFV, $OFV^{(i)}_{best} = f(P^{(i)})$ ($i = 1, \ldots, S$), and a \textit{global best} OFV, $OFV_{global}=f(P_{global})$.
    \end{itemize}
    \item the procedure is repeated until a \textit{stopping criterion} is met or the maximum number of iterations is reached. 
\end{enumerate}

\newpage
\bibliographystyle{unsrtnat}
\bibliography{template} 

\end{document}